\documentclass[12pt]{amsart}

\usepackage{epsfig}

\newcommand{\kahler}{K\"ahler\ }
\newcommand{\sqrtn}{\sqrt{N}}
\newcommand{\wt}{\widetilde}
\newcommand{\wh}{\widehat}
\newcommand{\PP}{{\mathbb P}}
\newcommand{\R}{{\mathbb R}}
\newcommand{\C}{{\mathbb C}}

\newcommand{\Z}{{\mathbb Z}}

\newcommand{\CP}{\C\PP}
\renewcommand{\d}{\partial}
\newcommand{\dbar}{\bar\partial}
\newcommand{\ddbar}{\partial\dbar}

\newcommand{\E}{{\mathbf E}\,}
\renewcommand{\H}{{\mathbf H}}
\newcommand{\half}{{\frac{1}{2}}}
\newcommand{\vol}{{\operatorname{Vol}}}
\newcommand{\codim}{{\operatorname{codim\,}}}
\newcommand{\SU}{{\operatorname{SU}}}
\newcommand{\FS}{{{\operatorname{FS}}}}

\renewcommand{\phi}{\varphi}
\newcommand{\eqd}{\buildrel {\operatorname{def}}\over =}
\newcommand{\nhat}{\raisebox{2pt}{$\wh{\ }$}}

\newcommand{\ccal}{\mathcal{C}}
\newcommand{\dcal}{\mathcal{D}}

\newcommand{\fcal}{\mathcal{F}}
\newcommand{\gcal}{\mathcal{G}}
\newcommand{\hcal}{\mathcal{H}}
\newcommand{\jcal}{\mathcal{J}}
\newcommand{\lcal}{\mathcal{L}}
\newcommand{\pcal}{\mathcal{P}}
\newcommand{\qcal}{\mathcal{Q}}
\newcommand{\ocal}{\mathcal{O}}
\newcommand{\scal}{\mathcal{S}}

\newcommand{\ucal}{\mathcal{U}}

\newcommand{\al}{\alpha}
\newcommand{\be}{\beta}
\newcommand{\ga}{\gamma}

\newcommand{\La}{\Lambda}
\newcommand{\la}{\lambda}
\newcommand{\ep}{\varepsilon}
\newcommand{\de}{\delta}
\newcommand{\De}{\Delta}
\newcommand{\om}{\omega}
\newcommand{\di}{\displaystyle}
\newcommand{\bbb}{|\!|\!|}
\newtheorem{theo}{{\sc Theorem}}[section]
\newtheorem{cor}[theo]{{\sc Corollary}}
\newtheorem{lem}[theo]{{\sc Lemma}}

\newenvironment{rem}{\medskip\noindent{\it Remark:\/} }{\medskip}
\newenvironment{defin}{\medskip\noindent{\it Definition:\/} }{\medskip}
\newenvironment{claim}{\medskip\noindent{\it Claim:\/} }{\medskip}

\title[Universality and scaling of correlations between zeros]
{Universality and scaling of correlations between zeros on complex
manifolds}

\author{Pavel Bleher}
\address{Department of Mathematical Sciences, IUPUI, Indianapolis, IN
46202,
USA}
\email{bleher@math.iupui.edu}
\author{Bernard Shiffman}
\address{Department of Mathematics, Johns Hopkins University, Baltimore,
MD
21218, USA} 
\email{shiffman@math.jhu.edu}
\author{Steve Zelditch}
\address{Department of Mathematics, Johns Hopkins University, Baltimore,
MD
21218, USA} 
\email{zel@math.jhu.edu}

\thanks{Research partially supported by NSF grants
\#DMS-9623214 (first author), \#DMS-9800479 (second author),
\#DMS-9703775
(third author).} 

\date{April 21, 1999}

\begin{document}

\begin{abstract} We study the limit as
$N\to\infty$ of the correlations between simultaneous zeros of random
sections of the powers $L^N$ of a positive holomorphic line bundle $L$
over a compact complex manifold $M$, when distances are rescaled so that
the average density of zeros is independent of $N$.  We show that the
limit
correlation is independent of the line bundle and depends only on the
dimension  of
$M$ and the codimension of the zero sets. We also provide some explicit
formulas for pair correlations. In particular, we provide an alternate
derivation of Hannay's limit pair correlation function for  $\SU(2)$
polynomials, and we show that this correlation function holds for all
compact Riemann surfaces.
\end{abstract}

\maketitle

\section*{Introduction}

This paper is concerned with  the local statistics of the simultaneous
zeros of $k$ random holomorphic sections $s_1, \dots, s_k   \in H^0(M,
L^N)$   of the $N^{\rm th}$ power $L^N$
of a positive Hermitian holomorphic line bundle $(L,h)$ over a compact
\kahler manifold $M$ (where $k\le m=\dim M$).  The terms `random' and
`statistics' are with respect to  a natural Gaussian probability measure
$d \nu_N$ on
$H^0(M, L^N)$ which we define below.    In the special case where
$M = \CP^m$ and $L$ is the
hyperplane section bundle $\mathcal{O}(1)$, sections of $L^N$ correspond
to holomorphic
polynomials of  degree $N$, and $(H^0(\CP^m, \mathcal{O}(N)), d \nu_N)$
is known as the
ensemble of $\SU(m+1)$ polynomials in the physics literature.
To obtain  local
statistics, we  expand a ball $U$ around a given point $z^0$ by a factor
$\sqrt N$ so that the average density of simultaneous scaled
zeros is independent of $N$.  We then ask whether the  simultaneous
scaled zeros 
behave as if thrown independently in $\sqrt{N} U$ or  how they are
correlated. 
Correlations between (unscaled) zeros are measured  by the so-called
{\it $n$-point zero
correlation function\/}  $K_{nk}^N(z^1,\dots,z^n)$, and those between
scaled zeros are
measured by the scaled correlation function $K_{nk}^N(\frac{z^1}{
\sqrtn},\dots,\frac{z^n}{ \sqrtn})$. 
  Our main result is that the large $N$
limits of the  scaled 
$n$-point correlation functions $K_{nk}^N(\frac{z^1}{
\sqrtn},\dots,\frac{z^n}{ \sqrtn})$  exist
and  are universal, i.e. are independent of $M,\ L$ and $h$ as well as
the
point $z^0$.  Moreover, the  scaling limit correlation functions can be
calculated explicitly.  We find that the limit correlations are short
range, i.e. that simultaneous scaled  zeros behave quite independently
for large distances.  On the other hand,
nearby  zeros exhibit some degree of repulsion.

To state our problems and results more precisely, we begin with provisional
definitions of the correlation functions $K_{nk}^N(z^1,\dots,z^n)$ and of the
scaling limit. (See \S\S \ref{notation}--\ref{corfuns} for the complete
definitions and notation.)  In order to provide a standard yardstick for our
universality results, we give $M$ the \kahler metric $\om$ given by the
(positive) curvature form of $h$.  The metrics $h$ and $\om$ then induce a
Hilbert space inner product on the space $H^0(M, L^N)$ of holomorphic sections
of $L^N$, for each $N\ge 1$.  In the spirit of \cite{SZ} we use this
$\lcal^2$-norm to define a Gaussian probability measure $d \nu_N$ on $H^0(M,
L^N)$. When we speak of a random section, we mean a section drawn at random
from this ensemble.  More generally, we can draw $k$ sections $(s_1, \dots,
s_k)$ independently and at random from this ensemble. Let $Z_{(s_1, \dots,
s_k)}$ denote their simultaneous zero set and let $|Z_{(s_1, \dots, s_k)}|$
denote the ``delta measure" with support on $Z_{(s_1, \dots, s_k)}$ and with
density given by the natural Riemannian volume $(2m - 2k)$-form defined by the
metric $\omega$.  To define the $n$-point zero correlation measure
$K_{nk}^N(z^1, \dots, z^n)$ we form the product measure
$$|Z_{(s_1, \dots, s_k)}|^n=\big(\underbrace{|Z_{(s_1, \dots,
s_k)}|\times\cdots\times|Z_{(s_1, \dots, s_k)}|}_{n}\big)\quad
\mbox{on}\quad
M^n:= \underbrace{M\times\cdots\times
M}_n\,.$$   To avoid trivial self-correlations, we puncture out the
generalized diagonal
in $M^n$ to get the punctured product space 
$$M_n=\{(z^1,\dots,z^n)\in M^n: z^p\ne z^q \ \ {\rm for} \ p\ne q\}\,.$$
We then restrict $|Z_{(s_1, \dots, s_k)}|^n$ to $M_n$ and define
$K_{nk}^N(z^1, \dots, z^n)$
to be the 
expected value $E(|Z_{(s_1, \dots, s_k)}|^n)$ of this measure with
respect to $\nu_N$.  When $k =
m$, the simultaneous
zeros almost surely form a discrete set of points and so this case is
perhaps the most vivid.  Roughly speaking,
$K_{nk}^N(z^1, \dots, z^n)$   gives the  probability density
of finding
simultaneous  zeros at 
$(z^{1}, \dots, z^n)$.

The first correlation function $K_{1kN}$ just gives the expected
distribution of simultaneous zeros of $k$ sections. 
In a previous paper
\cite{SZ} by two of the authors, it was shown (among other things) that
the expected
distribution of zeros is asymptotically uniform; i.e. 
$$K_{1k}^N(z^0)=c_{mk}N^k+O(N^{k-1})\,,$$ for any positive line bundle
(see
\cite[Prop.~4.4]{SZ}). The question then arises of
determining the
higher correlation functions.  As was first observed by \cite{BBL} and
\cite
{H} for $\SU(2)$ polynomials and by \cite{BD} for real polynomials in
one
variable, the zeros of a random polynomial are non-trivially correlated,
i.e. the zeros are not thrown down like independent points.  We will
prove
the same  for all  $\SU(m+1)$ polynomials and hence, by universality of
the scaling limit, for any  $M, L, h$.

To introduce the scaling limit, let us return to the case $k = m$ where
the simultaneous zeros form a discrete set of points. Since an
$m$-tuple of sections of $L^N$ will have $N^m$ times
as many zeros as $m$-tuples of sections of $L$, it is natural to
expand
$U$ by a factor of $\sqrtn$ to get a density of zeros that is
independent
of $N$. That is, we  choose  coordinates
$\{z_q\}$ for which
$z^0=0$ and
$\omega(z^0)=\frac{i}{2}\sum_q dz_q\wedge d\bar z_q$  
and then rescale
$z \mapsto \frac{z}{
\sqrtn}$.  Were the zeros thrown independently and at random on $U$, the
conditional probability density of finding a simultaneous zero at a
point
$w$ given a zero at $z$ would
be a constant independent of $(z,w)$.  Non-trivial correlations
(for any codimension $k\in\{1,\dots,m\}$) are measured by the
difference between $1$ and   the (normalized) {\it
$n$-point scaling limit zero correlation function\/}
$$\wt K_{nkm}^\infty(z^1,\dots,z^n) 
=\lim_{N\to\infty}\left(c_{mk}N^k\right)^{-n}
{K_{nk}^N\left(\frac{z^1}{ \sqrtn},\dots,\frac{z^n}{
\sqrtn}\right)}\,,\qquad
(z^1,\dots,z^n)\in U_n\,.$$

Our main result (Theorem \ref{uslc}) is universality of the scaling
limit
correlation functions:

\bigskip\begin{quote} 
{\it The $n$-point scaling limit zero correlation function
$\wt K_{nkm}^\infty(z^1,\dots,z^n)$ is given by a universal rational
function, homogeneous of degree $0$, in the values of the function
$e^{i\Im
(z\cdot \bar w)-\half |z-w|^2}$ and its first and second derivatives at
the
points
$(z,w)=(z^p,z^{p'})$, $1\le p,p'\le n$.  Alternately it is a
rational function in $z^p_q, \bar z^p_q, e^{z^p\cdot \bar z^{p'}}$}
\end{quote}

\bigskip The function $e^{i\Im (z\cdot \bar w)-\half |z-w|^2}$ which
appears
in the universal scaling limit is (up to a constant factor) the Szeg\"o
kernel
$\Pi_1^\H(z,w)$ of level
one for the reduced Heisenberg group ${\bf H}^n_{\rm red}$ (cf.\
\S \ref{notation}). Its appearance here owes to the fact that the
correlation
functions can be expressed in terms of the Szeg\"o kernels $\Pi_N(x,y)$
of
$L^N$. I.e., let $X$ denote the circle bundle over $M$ consisting of
unit
vectors in
$L^*$; then $\Pi_N(x,y)$ is the kernel of the orthogonal projection
$\Pi_N:\lcal^2(X)\to \hcal^2_N(X)
\approx H^0(M, L^N)$.  Indeed we have (Theorem \ref{npointcor}):

\bigskip
\begin{quote} {\it The $n$-point correlation $\wt
K_{nk}^N(z^1,\dots,z^n)$
is
given  by the above universal
rational function, applied this time to the values of the Szeg\"o kernel
$\Pi_N$ and its first and second derivatives at the points
$(z^p,z^{p'})$. }
\end{quote}

\bigskip
In view of this relation between the correlation functions and the
Szeg\"o
kernel, it suffices for the proof of the universality theorem \ref{uslc}
to
determine the scaling limit of the Szeg\"o kernel $\Pi_N$ and to show
its
universality. Indeed we shall show (Theorem \ref{neardiag}) that: 

\bigskip
\begin{quote} {\it Let $(z_1,\dots,z_m,\theta)$ denote local coordinates
in a neighborhood $\wt U\approx U\times S^1$ of a point $(z^0,\la)\in X$
(where  $(z_1,\dots,z_m)$ are the above local coordinates about
$z_0\in M$).  We then
have
$$N^{-m}\Pi_N\left(\frac{z}{\sqrtn},\frac{\theta}{N};
\frac{z'}{\sqrtn},\frac{\theta'}{N}\right)=
\Pi^\H_1(z,\theta;z',\theta') + O(N^{-1/2})\;.$$
}\end{quote}

\bigskip The fact that the correlation functions can be expressed
in terms of the Szeg\"o kernel may be explained in (at least) two
ways. The
first 
 is that the correlation functions may be expressed in terms of the
joint probability
density $D_{nk}^N(x,\xi;z)dxd\xi$ of the 
(vector-valued) random variable
$$(x,\xi)=\big[x^p,\xi^p\big]_{1\le p\le n}\;,\qquad
x^p=(s_1(z^p),\dots, s_k(z^p)),\quad \xi^p=
(\nabla s_1(z^p),
\dots, \nabla s_k(z^p))$$ given by the values
of the $k$
sections and of their covariant derivatives at the $n$ points $\{z^p\}$.
Our method of computing the correlation functions is based on the
following
probabilistic formula (Theorem \ref{density}):

\bigskip
\begin{quote} {\it  For $N$ sufficiently large so that the density
$D_{nk}^N
(x,\xi;z)$ is given by a continuous function, we have
$$
K_{nk}^N(z)=
\int  d\xi\,D_{nk}^N(0,\xi;z)\prod_{p=1}^n 
\det \left(\xi^{p}_j\xi^{p*}_{j'}\right)_{1\le j,j'\le k}\,,
\quad z=(z^1,\dots,z^n)\in M_n\,,$$
where $\xi=(\xi^1,\dots,\xi^n)$ 
and $\xi^{p*}_j:L^N_{z^p}\to T_{M,z^p}$ denotes
the adjoint to $\xi^p_j:T_{M,z^p}\to L^N_{z^p}$.}\end{quote}

\bigskip
This formula, which is valid in a more general setting, is based on
the approach of Kac \cite{Ka} and Rice \cite{Ri} (see also \cite {EK}) for
zeros of functions on $\R^1$, and of \cite{Halp} for zeros of (real) Gaussian
vector fields.  Since our probability measure $d\nu_N$ (on the space of
sections) is Gaussian, it follows that $D_{nk}^N$ is also a Gaussian density.
It will be proved in \S \ref{D-meet-S} that the covariance matrix of this
Gaussian may be expressed entirely in terms of $\Pi_N$ and its covariant
derivatives.  This type of formula for the correlation function of zeros was
previously used in \cite{BD}, \cite {H} and the works cited above.
We believe that this formula will have interesting applications in geometry.

A second
link between correlation functions and Szeg\"o kernels is given by the
Poincar\'e-Lelong
formula. In fact, this was our original approach to computing the
correlation
functions in the codimension 1 case. For the sake of brevity, we will
not
discuss this approach here; instead we refer the reader to our companion
article
\cite{BSZ}.

{From} the universality of our answers, it follows that the scaling
limit
pair
correlation functions depend only on the distance between points:
$$\wt K^\infty_{2km}(z^1,z^2)=\kappa_{km}(r)\,,\qquad r=|z^1-z^2|\,,$$
where $\kappa_{km}$ depends only on the dimension $m$ of $M$ and the
codimension $k$ of the zero set. In \S \ref{formulas}, we give explicit
formulas for the limit pair correlation functions $\kappa_{km}$
in some special cases.  
Our calculation uses the Heisenberg model, which (although
noncompact) is the
most natural one since   the scaled Szeg\"o kernels are all equal to
$\Pi_1$,
and there is no need in this case to take a limit. We also discuss the
hyperplane section bundle
$\ocal (1)
\to
\CP^m$, which is the most studied, since the sections of its powers are
the $\SU(m+1)$ polynomials---homogeneous polynomials in
$m+1$ variables---and the case $m=1$ (the $\SU(2)$ polynomials)
appears frequently in the physics literature (e.g., \cite{BBL, FH, H,
KMW,
PT}). We give expressions for the zero correlations $K^N_{nk}$ for the 
$\SU(m+1)$ polynomials and by letting $N\to\infty$, we obtain an
alternate derivation of our universal formula for the scaling limit
correlation.

We show (Theorem \ref{shortrange}) that $\kappa_{km}(r)=
1+O(r^4e^{-r^2})$ as $r\to +\infty$, and hence these correlations are
short
range in that they differ from the case of independent random points by
an
exponentially decaying term. We observe that when $\dim M=1$,
there is a strong repulsion between nearby zeros in the sense that 
$\kappa_{11}(r) \to 0$ as $r \to 0$, as was noted by Hannay \cite{H} and 
Bogomolny-Bohigas-Leboeuf
\cite{BBL} for the case of $\SU(2)$ polynomials. These asymptotics
are illustrated by the following graph (see also \cite{H}):

\vspace{-.5in}
\begin{figure}[ht]\label{kappa11}\centering
\epsfig{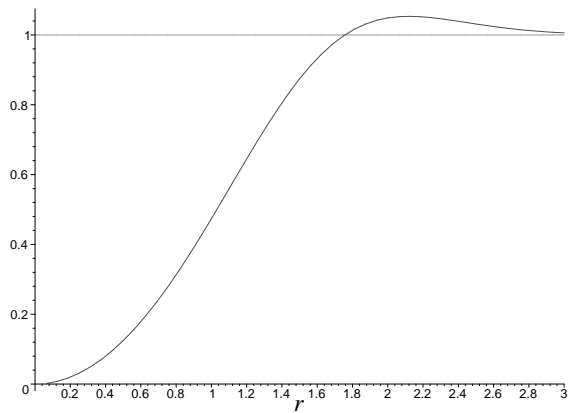}
\caption{The 1-dimensional limit pair correlation function
$\kappa_{11}$}
\end{figure}

For $\dim M=2$, the simultaneous scaled
zeros of a random pair $(s_1, s_2)$ of sections
still exhibit a mild  repulsion ($\lim_{r \to 0} \kappa_{22}(r) 
=\frac{3}{4}$), as illustrated in Figure~2 below.

\vspace{-.1in}
\begin{figure}[ht]\label{kappa22}\centering
\epsfig{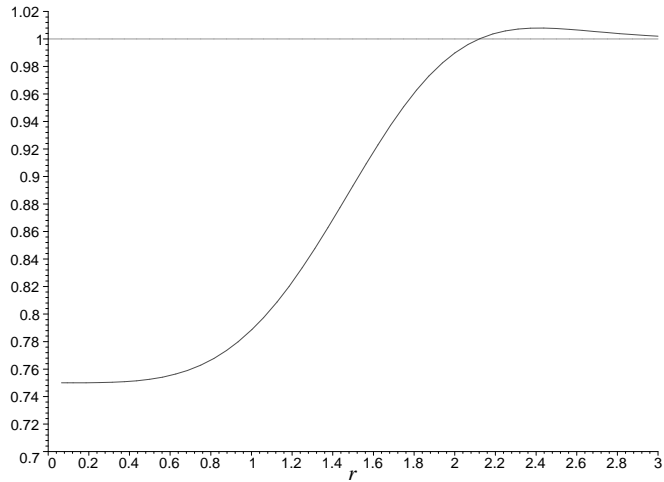}
\caption{The limit pair correlation function $\kappa_{22}$}
\end{figure}
The function $\kappa_{mm}(r)$ can be interpreted as the normalized
conditional probability of finding a zero near a point $z^1$ given that
there
is a zero at a second point a scaled distance $r$ from $z^1$ (in the
case of discrete zeros in $m$ dimensions).  The above graphs show that
for
dimensions 1 and 2, there is a unique scaled distance where this
probability
is maximized.  It would be interesting to explore the dependence of the
correlations on the dimension.  To ask one concrete question, do the
simultaneous scaled zeros in the point case become more and more
independent
in the sense that $\kappa_{mm} (r) \to 1$ as the dimension $m \to
\infty$?

When $k<m$, the zero sets are subvarieties of positive dimension $m-k$;
in
this case the expected volume of the zero set in a small spherical shell
of
radius $r$ and thickness $\ep$ about a point in the zero set must be
$\sim \ep
r^{2m-2k-1}$.  Hence we have $\kappa_{km}(r)\sim r^{-2k}$, for small
$r$.  The
graph of the limit correlation function for the case $m=2,k=1$ is given
in
Figure~3 below.

\vspace{-.25in}
\begin{figure}[ht]\label{kappa12}\centering
\epsfig{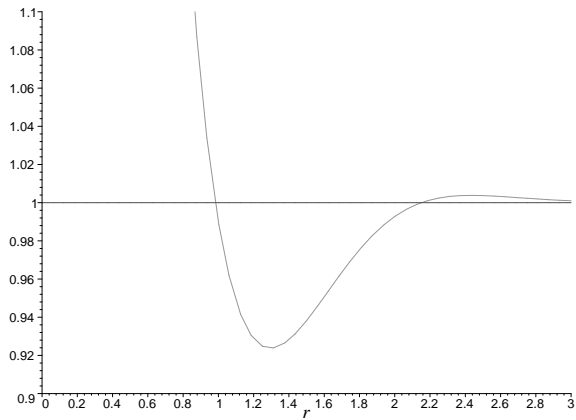}
\caption{The limit pair correlation function $\kappa_{12}$}
\end{figure}

To end this introduction, we would like to link our methods and results
at least
heuristically to a long tradition of (largely heuristic) results on
universality and scaling in statistical mechanics (cf. \cite{FFS}).  One
may
view the rescaling
transformation on $U$
as  generating a renormalization group.   The intuitive picture in
statistical
mechanics is that the renormalization
group should  carry a given system (read ``$L \to M$'') to the fixed
point
of the renormalization
group, i.e. to the scale invariant situation.  We  observe that the
local rescaling
of $U$ is nothing other than  the  Heisenberg dilations
$\delta_{\sqrt{N}}$ on 
${\bf H}^m_{\rm red}$. Since these dilations are automorphisms of the
(unreduced) Heisenberg group,
the Szeg\"o kernel of  ${\bf H}^m$ is invariant under  these dilations;
i.e., it  is the fixed point of the renormalization group. As predicted
by
this intuitive picture, we find that in the scaling limit all the
invariants of the line
bundle, in particular its zero-point 
correlation functions, are drawn to their values for the  fixed point
system
(read ``Heisenberg model'').

\bigskip
\section{Notation}\label{notation}

We begin with some notation and basic properties of sections of
holomorphic
line bundles, their zero sets,  Szeg\"o kernels, and Gaussian measures. 
We
also provide two examples that will serve as model cases for studying
correlations of zeros of sections of line bundles in the high power
limit.

\subsection{Sections of holomorphic line bundles}\label{cxgeom}

In this section, we introduce the basic complex analytic objects:
holomorphic
sections and the currents of integration over their zero sets.  We also
introduce Gaussian probability measures on spaces of holomorphic
sections.
For background in complex geometry,
we refer to \cite{GH}.

Let $M$ be a compact complex manifold and let $L\to M$ be a holomorphic line
bundle with a smooth Hermitian metric $h$; its curvature 2-form $\Theta_h$ is
given locally by
\begin{equation}\label{curvature}\Theta_h=-\ddbar
\log\|e_L\|_h^2\;,\end{equation} where $e_L$ denotes a local holomorphic
frame
(= nonvanishing section) of $L$ over an open set $U\subset M$, and
$\|e_L\|_h=h(e_L,e_L)^{1/2}$ denotes the $h$-norm of $e_L$.  
We say that $(L,h)$ is positive if the (real) 2-form
$\omega=\frac{\sqrt{-1}}{2}\Theta_h$ is positive, i.e., if $\om$ is a \kahler
form.  We henceforth assume that $(L,h)$ is positive, and
we give $M$
the
Hermitian metric corresponding to the \kahler form
$\omega$ and the induced Riemannian volume
form 
\begin{equation}\label{dV} dV_M= \frac{1}{m!}
\omega^m\;.\end{equation}
Since $\frac{1}{\pi}\om$ is a
de Rham representative of the Chern class $c_1(L)\in H^2(M,\R)$, the
volume of
$M$ equals $\frac{\pi^m}{m!}c_1(L)^m$.

The space $H^0(M, L^{N})$ of global holomorphic sections of
$L^N=L\otimes\cdots\otimes L$ is a finite dimensional complex vector space.
(Its dimension, given by the Riemann-Roch formula for large $N$, grows like
$N^m$.  By the Kodaira embedding theorem, the global sections of $L^N$ give an
embedding into a projective space for $N\gg 0$, and hence $M$ is a {\it
projective algebraic manifold.\/}) The metric $h$ induces Hermitian metrics
$h^N$ on $L^N$ given by $\|s^{\otimes N}\|_{h^N}=\|s\|_h^N$.  We give
$H^0(M,L^N)$ the Hermitian inner product
\begin{equation}\label{inner}\langle s_1, s_2 \rangle = \int_M h^N(s_1,
s_2)dV_M \quad\quad (s_1, s_2 \in H^0(M,L^N)\,)\;,\end{equation} and we
write
$|s|=\langle s,s \rangle^{1/2}$.

We now
explain our concept of a ``random section." We are
interested in expected values and correlations of zero sets of
$k$-tuples of
holomorphic sections of powers $L^N$.  Since the zeros do not depend on
constant factors, we could suppose our sections lie in the unit
sphere in
$H^0(M,L^N)$ with respect to the Hermitian inner product (\ref{inner}), 
and we pick random sections with respect to the spherical measure.
Equivalently, we could suppose that $s$ is a random element of the
projectivization
$\PP
H^0(M,L^N)$. Another equivalent approach is to use Gaussian measures on
the
entire space $H^0(M,L^N)$.  We shall use the third approach, since
Gaussian
measures seem the best for calculations.  Precisely, we give
$H^0(M,L^N)$ the complex
Gaussian probability measure
\begin{equation}\label{gaussian}d\nu_N(s)=\frac{1}{\pi^m}e^
{-|c|^2}dc\,,\qquad s=\sum_{j=1}^{d_N}c_jS_j^N\,,\end{equation} where
$\{S_j^N\}$ is an orthonormal basis for
$H^0(M,L^N)$ and $dc$ is $2d_N$-dimensional Lebesgue measure.  This
Gaussian
is characterized by the property that the $2d_N$ real variables $\Re
c_j,
\Im c_j$ ($j=1,\dots,d_N$) are independent random variables with mean 0
and
variance $\half$; i.e.,
$$\E c_j = 0,\quad \E c_j c_k = 0,\quad  \E c_j \bar c_k = \de_{jk}\,.$$
Here and throughout this paper, $\E$ denotes expectation.

In general, a {\it complex Gaussian measure\/} (with mean 0) on a finite
dimensional complex vector space $V$ is a measure $\nu$ of the form
(\ref{gaussian}), where the $c_j$ are the coordinates with respect to
some
basis. Explicitly, the complex Gaussian measures on $\C^m$ are the
probability measures of the form
\begin{equation}\label{gaussian2} \frac{e^{-\langle
\De^{-1}z,z\rangle}}{\pi^m\det
\De}dz \end{equation} where $\De=(\De^j_k)$ is a
positive definite Hermitian matrix and
$$\langle \zeta, z\rangle=\zeta\cdot\bar z=\sum_{q=1}^m \zeta_q\bar
z_q$$
denotes the standard Hermitian inner product in $\C^m$.  For the
Gaussian
measure (\ref{gaussian2}), we have
\begin{equation}\label{gaussian3} \E(z_j z_k)=0, \qquad  \E(z_j \bar
z_k)=\De^j_k\,.\end{equation} 
If $\nu$ is a complex Gaussian on $V$ and
$\tau:V\to
\wt V$ is a surjective linear transformation, then $\tau_*\nu$ is a
complex
Gaussian on $\wt V$. In particular, if $\wt V=\C^m$, then, $\tau_*\nu$
is of
the form (\ref{gaussian2}), where the covariance matrix $\De$ is given
by
(\ref{gaussian3}) with $z_j=z_j\circ\tau:V\to\C$. 

We shall consider the space $\scal=H^0(M,L^N)^k$ ($1\le k \le m$) with
the
probability measure $d\mu=d \nu\times\cdots\times d\nu$, which is also
Gaussian.  Picking a random element of $\scal$ means picking $k$
sections of
$H^0(M,L^N)$ independently and at random.  For
$s=(s_1,\dots,s_k)\in\scal$, we
let
$$Z_s=\{z\in M: s_1(z)=\cdots=s_k(z)=0\}$$
denote the zero set of $s$.
Note that if $N$ is sufficiently large so that $L^N$ is base point free,
then for
$\mu$-a.a.
$s\in\scal$, we have
$\codim Z_s=k$.  (Indeed,   the set of $s$
where
$\codim Z_s<k$ is a proper algebraic subvariety of $H^0(M,L^N)^k$.  In
fact, by Bertini's theorem, the $Z_s$ are smooth submanifolds of complex
dimension $m-k$ for almost all $s$, provided $N$ is large enough so that
the
global sections of $L^N$ give a projective embedding of $M$, but we do
not
need this fact here.)  For these $s$, we let
$|Z_s|$ denote Riemannian 
$(2m-2k)$-volume along the regular points of $Z_s$, regarded as a
measure on
$M$: 
\begin{equation}\label{volmeasure} (|Z_s|,\phi)=\int_{Z_s^{\rm reg}}\phi
d\vol_{2m-2k}=\frac{1}{(m-k)!}\int_{Z_s^{\rm reg}}\phi
\omega^{m-k}\,.\end{equation} It was shown by Lelong \cite{Le}
(see also \cite{GH}) that
the integral in (\ref{volmeasure}) converges.
(In fact, $|Z_s|$ can be regarded as the total variation measure of 
the closed current of integration
over $Z_s$.)
We regard $|Z_s|$ as a measure-valued random variable on the
probability space $(\scal,d\mu)$; i.e., for each test function
$\phi\in\ccal^0(M)$, $(|Z_s|,\phi)$ is a complex-valued random variable.

\subsection{Szeg\"o kernels}\label{Szegokernels}

As in \cite{Z,SZ} we now lift the analysis of holomorphic sections
over $M$ to 
a certain $S^1$ bundle $X \to M$.  This is a useful approach to the
asymptotics of powers
of line bundles and goes back at least to  \cite{BG}.  

We let $L^*$ denote the dual line bundle to $L$, and we consider the
circle
bundle $X=\{\la \in L^* : \|\la\|_{h^*}= 1\}$, where $h^*$ is the norm
on
$L^*$ dual to $h$. Let $\pi:X\to M$ denote the bundle map; if $v\in
L_z$, then
$\|v\|_h=|(\la,v)|$, $\la\in X_z=\pi^{-1}(z)$. Note that $X$ is the
boundary
of the
disc bundle $D = \{\la \in L^* : \rho(\la)>0\}$, where
$\rho(\la)=1-\|\la\|^2_{h^*}$. The disc bundle $D$ is strictly
pseudoconvex in
$L^*$, since $\Theta_h$ is positive, and hence $X$ inherits the
structure of
a strictly pseudoconvex CR manifold.  Associated to $X$ is the contact
form
$\al= -i\partial\rho|_X=i\dbar\rho|_X$.  We also give $X$ the volume
form
\begin{equation}\label{dvx}dV_X=\frac{1}{m!}\al\wedge 
(d\al)^m=\al\wedge\pi^*dV_M\,.\end{equation}

The setting for our analysis of the Szeg\"o kernel is the Hardy space
$\hcal^2(X)
\subset \lcal^2(X)$ of square-integrable CR functions on $X$, i.e.,
functions that are annihilated by the
Cauchy-Riemann operator $\dbar_b$ (see \cite[pp.~592--594]{Stein}) and
are
$\lcal^2$ with respect to the inner product
\begin{equation}\label{unitary} \langle  F_1, F_2\rangle
=\frac{1}{2\pi}\int_X
F_1\overline{F_2}dV_X\,,\quad F_1,F_2\in\lcal^2(X)\,.\end{equation}
Equivalently, $\hcal^2(X)$
is the space of boundary values of holomorphic functions on $D$ that
are
in
$\lcal^2(X)$.  We let $r_{\theta}x =e^{i\theta} x$ ($x\in X$) denote the
$S^1$
action on $X$ and denote its infinitesimal generator by
$\frac{\partial}{\partial \theta}$. The $S^1$ action on $X$ commutes
with $\bar{\partial}_b$; hence $\hcal^2(X) = \bigoplus_{N
=0}^{\infty} \hcal^2_N(X)$ where $\hcal^2_N(X) = 
\{ F \in \hcal^2(X): F(r_{\theta}x)
= e^{i
N \theta} F(x) \}$. A section $s$ of $L$ determines an equivariant
function
$\hat{s}$ on $L^*$ by the rule $\hat{s}(\lambda) = \left(\lambda,
s(z)
\right)$ ($\lambda \in L^*_z, z \in M$). It is clear that if $\tau \in
\C$
then $\hat{s}(z, \tau \lambda) = \tau \hat{s}$. We henceforth restrict
$\hat{s}$ to $X$ and then the equivariance property takes the form
$\hat{s}(r_{\theta} x) = e^{i \theta}\hat{s}(x)$.  Similarly, a section
$s_N$
of $L^{N}$ determines an equivariant function $\hat{s}_N$ on $X$: put
$$\hat{s}_N(\lambda) = \left( \lambda^{\otimes N}, s_N(z)
\right)\,,\quad
\la\in X_z\,,$$ where $\lambda^{\otimes N} = \lambda \otimes
\cdots\otimes
\lambda$;
then $\hat s_N(r_\theta x) = e^{iN\theta} \hat s_N(x)$.  The map
$s\mapsto
\hat{s}$ is a unitary equivalence between $H^0(M, L^{ N})$ and
$\hcal^2_N(X)$. (This follows from (\ref{dvx})--(\ref{unitary}) and the
fact
that
$\alpha= d\theta$ along the fibers of $\pi:X\to M$.)

We let $\Pi_N : \lcal^2(X) \rightarrow \hcal^2_N(X)$ denote the
orthogonal
projection.  The Szeg\"o kernel $\Pi_N(x,y)$ is defined by
\begin{equation} \Pi_N F(x) = \int_X \Pi_N(x,y) F(y) dV_X (y)\,,
\quad F\in\lcal^2(X)\,.
\end{equation} It can be given as
\begin{equation}\label{szego}\Pi_N(x,y)=\sum_{j=1}^{d_N}\wh
S_j^N(x)\overline{\wh S_j^N(y)}\,,\end{equation} where
$S_1^N,\dots,S_{d_N}^N$ form an orthonormal basis of $H^0(M,L^N)$.  Pick
a
local holomorphic frame $e_L$ for $L$ over an open subset $U\subset M$,
let
$e_L^*$ denote the dual frame, and write
$h(z)=h(e_L(z),e_L(z))=\|e_L\|_h^2$. 
The map
$(z,e^{i\theta})\mapsto e^{i\theta}h(z)^{1/2} e_L^*(z)$ gives an
isomorphism
$U\times S^1\approx \pi^{-1}(U)\subset X$, and we use the coordinates
$(z,\theta)$ to identify points of $\pi^{-1}(U)$. For $s\in H^0(M,L^N)$,
we have
\begin{equation} \label{hats}\hat s(z,\theta)=
\left\langle s(z),e^{iN\theta}h(z)^{N/2}e^*_L
(z)\right\rangle = e^{iN\theta}h(z)^{N/2} f(z),\quad s=fe_L^{\otimes
N}\,.
\end{equation}

Although the Szeg\"o kernel is defined on $X$, its absolute value is
well-defined on $M$ as follows: writing $S_j^N=f^N_j e_L^{\otimes N}$,
we
have
\begin{equation}\label{PiN}\Pi_N(z,\theta;w,\phi)
=e^{iN(\theta-\phi)}\Pi_N(z,0;w,0)= e^{iN(\theta-\phi)}h(z)^{N/2}
h(w)^{N/2}\sum_{j=1}^{d_N}f^N_j(z)\overline{f^N_j(w)}\,,\end{equation}
for
$z,w\in U$. (Here we may take $U$ to be the disjoint union of connected
neighborhoods of $z$ and $w$, if $z$ is not close to $w$.)  Thus we can
write
$$|\Pi_N(z,w)|= |\Pi_N(z,0;w,0)|\,,$$ which is independent of the choice
of
local frame $e_L$.  On the diagonal we have $$\Pi_N(z,z)=
\Pi_N(z,\theta;z,\theta)=\sum_{j=1}^{d_N}\|S_j^N(z)\|_{h^N}\,.$$ 

The
Hermitian
connection $\nabla$ on $L$ induces the decomposition $T_X=T_X^{H}\oplus
T_X^{V}$ into horizontal and vertical components, and
we let $t^H$ denote the horizontal lift (to $X$) of a vector field $t$
in $M$.
We consider the horizontal operators on $X$:
$$d_{z_q}^H \eqd d_{(\d /\d z_q)^H}\,,\quad 
d_{\bar z_q}^H\eqd d_{(\d /\d \bar z_q)^H}\,,$$
where $z_1,\dots,z_m,\theta$ denote local coordinates on
$X$.  We note that
\begin{equation} \label{horizontal} d_{z_q}^H \hat s = (\nabla^N_{z_q}
s)\nhat
\;,\quad s\in H^0(M,L^N)\,,\end{equation}
where $\nabla^N$ is the induced connection on $L^N$. We then have

\begin{eqnarray}d^H_{z_q}\Pi_N(z,\theta;w,\phi)
&=&\sum_{j=1}^{d_N}\left(\nabla^N_{z_q}
S_j^N\right)\nhat(x)\overline{\wh S_j^N(y)}\nonumber\\
&=&e^{iN(\theta-\phi)}h(z)^{N/2}
h(w)^{N/2}\sum_{j=1}^{d_N}f^N_{j;q}(z)\overline{f^N_j(w)}\,,\nonumber\\
\label{dPiN} d^H_{z_p} d^H_{\bar w_q}\Pi_N(z,\theta;w,\phi)&=&
\sum_{j=1}^{d_N}\left(\nabla^N_{z_p}
S_j^N\right)\nhat(x)\overline{\left(\nabla^N_{w_q}
S_j^N\right)\nhat(y)}\\
&=&e^{iN(\theta-\phi)}h(z)^{N/2}
h(w)^{N/2}\sum_{j=1}^{d_N}f^N_{j;p}(z)\overline{f^N_{j;q}(w)}\,,
\nonumber
\end{eqnarray} $$ \qquad \nabla^N_{z_q}=\nabla^N_{\d /\d z_q}\,,\quad
f^N_{j;q} =\frac{\d f}{\d z_q}+Nf^N_jh^{-1}\frac{\d
h}{\d z_q}\,.$$
We can also use (\ref{hats}) and (\ref{horizontal}) to describe the
horizontal
lift in local coordinates:
\begin{equation}\label{horizontal2} d^H_{z_q}= \frac{\d}{\d z_q}
-\frac{i}{2}
\frac{\d\log h}{\d z_q}\frac{\d}{\d\theta}\,.\end{equation}

\subsection{Model examples}  In two special cases we can work out the
Szeg\"o
kernels and their derivatives
explicitly, namely for the hyperplane section bundle over  $\CP^m$ and
for
the Heisenberg bundle over $\C^m$, i.e.
the trivial line bundle with curvature equal to the standard symplectic
form on $\C^m$. 
These cases will be
important after   we have proven universality, since scaling limits of
correlation functions
for all line bundles coincide with those of the  model cases.

In fact, the two models are locally  equivalent in the CR sense.
In the case of
$\CP^m$, the circle bundle $X$ is the $2m + 1$ sphere $S^{2m +
1}$, which is the boundary of the unit ball $B^{2m + 2} \subset
\C^{m+1}$. In the case of $\C^m$, the
circle bundle is the reduced Heisenberg group ${\bf H}^m_{\rm red}$,
which
is a discrete quotient of the simply connected
Heisenberg group $\C^m \times \R$. As is well-known, the latter is
equivalent (in the CR and contact sense) to the
boundary of $B^{2m + 2} $ (\cite{Stein}).

\subsubsection{$\SU(m+1)$-polynomials}\label{example1}
For our first example, we let
$M=\CP^m$ and take
$L$  to be the hyperplane section bundle $\ocal(1)$. 
Sections $s\in
H^0(\C\PP^m,\ocal(1))$ are linear functions on
$\C^{m+1}$; the zero divisors $Z_s$ are projective hyperplanes. The line
bundle $\ocal(1)$ carries a natural metric $h_\FS$ given by
\begin{equation}\label{hfs} \|s\|_{h_\FS}([w])=\frac{|(s,w)|}{|w|}\;, 
\quad\quad
w=(w_0,\dots,w_m)\in\C^{m+1}\;,\end{equation} for $s\in\C^{m+1*}\equiv
H^0(\C\PP^m,\ocal(1))$, where $|w|^2=\sum_{j=0}^m |w_j|^2$ and
$[w]\in\C\PP^m$
is the complex line through $w$. The \kahler form on $\CP^m$ is the
Fubini-Study form
\begin{equation}
\omega_\FS=\frac{\sqrt{-1}}{2}\Theta_{h_\FS}=\frac{\sqrt{-1}}{2}
\ddbar \log |w|^2 \,.\end{equation}
The dual bundle $L^*=\ocal(-1)$ is the affine space
$\C^{m+1}$ with the origin blown up, and $X=S^{2m+1}\subset\C^{m+1}$.
The
$N$-th tensor power of $\ocal(1)$ is denoted $\ocal(N)$.  Elements
$s_N\in
H^0(\C\PP^m,\ocal(N))$ are homogeneous polynomials on $\C^{m+1}$ of
degree
$N$, and $\hat s_N=s_N|_{S^{2m-1}}$.  The monomials
\begin{equation}\label{orthonormal}s^N_J =
\left[\frac{(N+m)!}{\pi^m j_0!\cdots j_m!}\right]^\half z^J\,,\quad z^J=
z_0^{j_0}\cdots z_m^{j_m},\quad\quad J=(j_0,\ldots,j_m),\
|J|=N\end{equation}
form an orthonormal basis for $H^0(\C\PP^m,\ocal(N))$. (See \cite[\S
4.2]{SZ};
the extra factor $\left(\frac{m!}{\pi^m}\right)^{1/2}$ in
(\ref{orthonormal})
comes from the fact that here $\CP^m$ has the usual volume
$\frac{\pi^m}{m!}$, whereas in \cite{SZ}, the volume of $\CP^m$
is normalized to be 1.) Hence the Szeg\"o
kernel for $\ocal(N)$ is given by
\begin{equation}\label{Szegosphere} \Pi_N(x,y)=\sum_J
\frac{(N+m)!}{\pi^mj_0!\cdots j_m!}x^J \bar y^J =  \frac{(N+m)!}
{\pi^mN!}\langle x,y\rangle^N\,.\end{equation}
Note that
$$\Pi(x,y)=\sum_{N=1}^\infty\Pi_N(x,y)=\frac{m!}{\pi^m}
(1-\langle x, y\rangle)^{-(m+1)} =
2\pi\times [\mbox{classical Szeg\"o kernel on} \
S^{2m+1}]\,.$$ (The factor
$2\pi$ is due to our normalization (\ref{unitary}).)

\subsubsection{The Heisenberg model}\label{example2} 
Our second
example is  the linear model $\C^m
\times \C \to \C^m$  for  positive line bundles $L \to M$ over \kahler
manifolds and their associated Szeg\"o kernels.  It is most illuminating
to consider the associated
principal $S^1$ bundle $\C^m \times S^1 \to \C^m$, which may be
identified with the boundary of the disc bundle $D \subset L^*$ in the
dual line bundle.  This $S^1$ bundle is the {\it
reduced Heisenberg group} ${\bf H}_{\rm red}^m$ (cf. \cite{F}, p. 23). 

Let us recall its definition and properties. We start  with the usual
(simply connected) Heisenberg group ${\bf
H}^m$  (cf. \cite{F} \cite{Stein}; note that different authors differ by
factors of $2$ and $\pi$
in various definitions).  It is the 
group  $\C^m \times \R$ with  group law 
$$(\zeta, t) \cdot (\eta, s) = (\zeta + \eta, t + s +  \Im (\zeta
\cdot
\bar{\eta})).$$
The identity element is $(0, 0)$ and $(\zeta, t)^{-1} = (- \zeta, - t)$.
Abstractly, the  Lie algebra of ${\bf H}_m$ is spanned by elements $Z_1,
\dots, Z_m, \bar{Z}_1, \dots, \bar{Z}_m, T$ satisfying the canonical
commutation relations $[Z_j, \bar{Z}_k] = -i  \delta_{jk}  T$ (all other
brackets zero). Below
we will select such a basis of left invariant vector fields. 

${\bf H}^m$ is a strictly convex CR manifold
which may be embedded in $\C^{m + 1}$ as the boundary of a strictly
pseudoconvex domain, namely  
the upper half space $ \ucal ^m :=  \{z \in \C^{m + 1}: \Im z_{m + 1}
> \half  \sum_{j = 1}^m |z_j|^2 \}$.
The boundary of $\ucal ^m$ equals $\partial \ucal ^m = \{z \in \C^{m
+ 1}: \Im z_{m + 1} = \half
 \sum_{j = 1}^m |z_j|^2\}$.
${\bf H}^m$ acts simply transitively on $\partial \ucal ^m$ (cf.
\cite{Stein}, XII), and we get 
an identification of ${\bf H}^m$ with $\partial \ucal ^m$ by:
$$[\zeta, t] \to (\zeta, t + i |\zeta|^2) \in \partial \ucal ^m.$$

The Szeg\"o projector of ${\bf H}^m$ is the operator $\Pi: \lcal^2({\bf
H}^m) \to
\hcal^2({\bf H}^m)$ 
of orthogonal projection onto boundary values of holomorphic functions
on
$\ucal ^m$ which lie in $\lcal^2$. The kernel of
$\Pi$ is given by (cf. \cite{Stein}, XII \S 2 (29))
$$\Pi(x,y) = K(y^{-1} x), \;\;\;\;\;\;\; K(x) = - C_m
\frac{\partial}{\partial t} [t + i |\zeta|^2]^{-m} \in\dcal'(\H^m)\,.$$

The linear model for the principal $S^1$ bundle described in \S 1.2 is
the
so-called reduced Heisenberg group ${\bf H}^m_{\rm red}={\bf H}^m/ \{(0,
2\pi
k): k \in \Z\} = \C^m \times S^1$ with group law $$(\zeta, e^{i t})
\cdot
(\eta, e^{i s}) = (\zeta + \eta, e^{i(t + s + \Im (\zeta \cdot
\bar{\eta}))}).$$ It is the principal $S^1$ bundle over $\C^m$
associated to
the line bundle $L_\H=\C^m\times\C$.  The metric on $L_\H$ with
curvature
$\Theta=\ddbar |z|^2$ is given by setting $h_\H(z)=e^{-|z|^2}$; i.e.,
$|f|_{h_\H}= |f|e^{-|z|^2/2}$. The reduced Heisenberg group ${\bf
H}^m_{\rm
red}$ may be viewed as the boundary of the dual disc bundle $D \subset
L^*_\H$
and hence is a strictly pseudoconvex CR manifold.

It seems most natural to approach the analysis of the Szeg\"o kernels on
${\bf
H}^m_{\rm red}$ from the representation-theoretic point of view.  Let us
begin
with the case $N = 1$.  We thus consider the space $\mathcal{V}_1
\subset
\lcal^2({\bf H}^m_{\rm red})$ of functions $f$ satisfying
$\frac{1}{i}\frac{\partial}{ \partial \theta} f = f$, which forms a
(reducible) representation of ${\bf H}^m_{\rm red}$ with central
character
$e^{i \theta}$.  By the Stone-von Neumann theorem there exists a unique
(up to
equivalence) representation $(V_1, \rho_1)$ with this character and by
the
Plancherel theorem, $\mathcal{V}_1 \cong V_1 \otimes V_1^*$.

The space of CR functions in $\mathcal{V}_1$ is an irreducible invariant
subspace.  Here, by CR functions we mean the functions satisfying the
left-invariant Cauchy-Riemann equations $\bar{Z}^L_q f = 0$ on ${\bf
H}^m_{\rm
red}$.  Here, $\{\bar{Z}^L_q\}$ denotes a basis of the left-invariant
anti-holomorphic vector fields on ${\bf H}^m_{\rm red}$.  Let us recall
their
definition: we first equip ${\bf H}^m_{\rm red}$ with its left-invariant
contact form $\alpha^L =  \sum_q(u_qdv_q-v_qdu_q) +
d\theta$ ($\zeta=u+iv$).  The left-invariant CR
holomorphic
(resp.\ anti-holomorphic) vector fields $Z_q^L$ (resp.\ $ \bar{Z}_q^L$)
are the
horizontal lifts of the vector fields $\frac{\partial}{\partial z_q}$
(resp.\
$\frac{\partial}{\partial \bar{z}_q}$) with respect to $\alpha^L$. They span
the
left-invariant CR structure of ${\bf H}^m_{\rm red}$ and the $Z_q^L$
obviously
have the form $Z_q^L = \frac{\partial}{\partial z_q} + A
\frac{\partial}{\partial \theta}$ where the coefficient $A$ is
determined by
the condition $\alpha^L(Z_q^L ) = 0$.  An easy calculation gives:
$$Z_q^L = \frac{\partial}{\partial z_q} + \frac{i}{2}\bar{z}_q
\frac{\partial}{\partial \theta}, \;\;\;\; \bar{Z}^L_q =
\frac{\partial}{\partial \bar{z}_q} - \frac{i}{2} z_q
\frac{\partial}{\partial
\theta}.$$ The vector fields $\{\frac{\partial}{\partial \theta}, Z_q^L,
\bar{Z}_q^L\}$ span the Lie algebra of ${\bf H}^m_{\rm red}$ and satisfy
the
canonical commutation relations above.

We then define the Hardy space $\mathcal{H}^2 ({\bf H}^m_{\rm red})$ of
CR
holomorphic functions, i.e. solutions of $\bar{Z}_q^L f = 0$, which lie
in
$\lcal^2({\bf H}^m_{\rm red})$. We also put $\mathcal{H}^2_1 =
\mathcal{V}_1
\cap \mathcal{H}^2({\bf H}^m_{\rm red}).$ The group ${\bf H}^m_{\rm
red}$ acts
by left translation on $\mathcal{H}^2_1$.  The generators of this
representation are the right-invariant vector fields $Z_q^R,
\bar{Z}_q^R$
together with $\frac{\partial}{\partial \theta}$. They are horizontal
with
respect to the right-invariant contact form $\alpha^R =
\sum_q(u_qdv_q-v_qdu_q) - d\theta$ and are given by:
$$Z_q^R = \frac{\partial}{\partial z_q} - \frac{i}{2}\bar{z}_q
\frac{\partial}{\partial \theta}, \;\;\;\; \bar{Z}^R_q =
\frac{\partial}{\partial \bar{z}_q} + \frac{i}{2} z_q
\frac{\partial}{\partial
\theta}\,.$$ In physics terminology, $Z_q^R$ is known as an annihilation
operator and $\bar{Z}_q^R$ is a creation operator.

The representation $\mathcal{H}^2_1$ is irreducible and may be identified with
the Bargmann-Fock space of entire holomorphic functions on $\C^m$ which are
square integrable relative to $e^{-|z|^2}$ (or equivalently, holomorphic
sections of the trivial line bundle $L_\H=\C^m\times \C$ mentioned above, with
hermitian metric $h_\H=e^{-|z|^2}$).  The identification goes as follows: the
function $\phi_0(z, \theta) := e^{i \theta} e^{-|z|^2/2}$ is CR holomorphic
and is also the ground state for the right invariant ``annihilation operator;''
i.e., it satisfies
$$\bar{Z}^L_q \phi_0(z, \theta) = 0 = Z_q^R \phi_0(z, \theta)\,.$$ Any element
$F(z,\theta)$ of $\mathcal{ H}^2_1$ may be written in the form $F (z, \theta)
= f(z) \phi_0.$ Then $\bar{Z}^L_q F =(\frac{\partial}{\partial \bar{z}_q} f )
\phi_0$, so that $F$ is CR if and only if $f$ is holomorphic. Moreover, $F \in
\lcal^2({\bf H}^m_{\rm red})$ if and only if $f$ is square integrable
relative to $e^{-|z|^2}$.

The Szeg\"o kernel $\Pi_1^\H(z, \theta, w, \phi)$ of ${\bf H}^m_{\rm
red}$ is by
definition the orthogonal projection from $\lcal({\bf H}^m_{\rm red})$
to
$H^2_1.$ As will be seen below, $\Pi_1^\H(z, \theta, w, \phi) = 
\frac{1}{\pi_m}e^{i (\theta
-\phi )} e^{ (z\cdot\bar w -\half |z|^2 -\half|w|^2) }$, which is the
left
translate of $\phi_0$ by $(-w, - \phi)$.  In the physics terminology it
is the
coherent state associated to the phase space point $w.$

So far we have set $N = 1$, but the story is very similar for any $N$.
We
define $\hcal^2_N$ as the space of square- integrable CR functions
transforming by $e^{i N \theta}$ under the central $S^1$.  By the
Stone-von
Neumann theorem there is a unique irreducible $V_N$ with this central
character. The main difference to the case $N = 1$ is that $\hcal^2_N$
is of
multiplicity $N^m$.  The Szeg\"o kernel $\Pi_N^\H(x,y)$ is the
orthogonal
projection to $\hcal^2_N$ and is given by the dilate of $\Pi^\H_1$. 
Thus,
$$\Pi_N^\H(x,y) =\frac{1}{\pi^m} N^m e^{i N (\theta -\phi )} e^{
N(z\cdot\bar
w -\half |z|^2 -\half|w|^2) }.$$

To prove these formulae for the Szeg\"o kernels, we observe that the
reduced
Szeg\"o kernels are obtained by projecting the Szeg\"o kernel on ${\bf
H}^m$ to
${\bf H}_{\rm red}^m$ as an automorphic kernel, i.e.  $$\Pi^\H(x, y) =
\sum_{n
\in \Z } \Pi(x, y\cdot (0, 2\pi n)).$$ Let us write $x = (z, \theta), y
= (w,
\phi)$.  Then the $N$-th Fourier component $\Pi^\H_N(x, y)$ of $\Pi^\H$,
i.e.
the projection onto square integrable holomorphic sections of $L^N$, is
given
by: \begin{eqnarray*}\Pi^\H_N(x, y)&=& \int_{\R} e^{- i N t} \Pi( e^{i
t} x, y
) dt\ =\ \int_{\R} e^{- i N t} K(e^{ i t} y^{-1} x ) dt\\&=& \int_{\R}
e^{- i
N t} K(z - w, e^{i (\theta - \phi + t + \Im (z \cdot \bar{w})} )
dt\,.\end{eqnarray*} Here we abbreviated the element $(0, e^{it})$ by
$e^{it}$.  Change variables $t \mapsto t- \theta + \phi - \Im (z \cdot
\bar{w})$ to get $$\begin{array}{l}\Pi^\H_N(x, y) = e^{i N (\theta -
\phi)}
e^{i N \Im (z \cdot \bar{w}}) \int_{\R} e^{- i N t} K(z - w, t) dt \\ \\
=
e^{i N ( \theta - \phi)}e^{i N \Im (z \cdot \bar{w})} \hat{K}_t(z - w,
N)\end{array} $$ where $\hat{K}_t$ is the Fourier transform of $K$ with
respect to the $t$ variable.  By \cite[p.~585]{Stein}, the full $\R^{2m}
\times \R$ Fourier transform of $K$ is given by $\hat{K} (z, N) = C'_m
e^{-|z|^2/ 2N}$, so by taking the inverse Fourier transform in the $z$
variable we get the Fourier transform just in the $t$ variable:
\begin{equation}\label{PiHN}  \Pi^\H_N(x, y) =
\frac{1}{\pi^m} N^m e^{i N (\theta -
\phi)}e^{  iN \Im (z \cdot \bar{w})} e^{- \half N |z -
w|^2}. \end{equation}
(Our constant factor $\frac{1}{\pi^m}$ in (\ref{PiHN}) is determined by
the
condition that $\Pi_N^\H$ is an orthogonal projection.)

In our study of the correlation functions, we will need explicit
formulae for
the horizontal derivatives of the Szeg\"o kernel.  The left-invariant
derivatives are given by
\begin{eqnarray} N^{-m} Z^L_q \Pi_N^\H(z,\theta;w,\phi)
&=&\, N (\bar{w}_q - \bar{z}_q) \Pi_N^\H(z,\theta;w,\phi)\,, \nonumber\\
\label{dPiNHleft} & & \\
N^{-m} Z^L_q  \bar{W}^L_{ q'} \Pi_N^\H(z,\theta;w,\phi)
&=& N^2 (z_{q'} - w_{q'})(\bar w_q - \bar{z}_q)
\Pi_N^\H(z,\theta;w,\phi) + N\de_{qq'}
\Pi_N^\H(z,\theta;w,\phi)\,.   \nonumber
\end{eqnarray}  
Comparing the definitions of the horizontal vector fields with
(\ref{horizontal}), using $h_\H=e^{-|z|^2}$, we see that $d^H_{z_q}=Z^L_q$, as
expected, since $\al^L$ agrees with the contact form $\al$ for $L_\H$ (as
defined in \S \ref{Szegokernels}). 
We
will see later that our formulas for computing correlations are valid
with any
connection, and thus it is sometimes useful to also consider the right
invariant derivatives:
\begin{eqnarray} N^{-m} Z^R_q \Pi_N^\H(z,\theta;w,\phi)
&=&\, N \bar{w}_q \Pi_N^\H(z,\theta;w,\phi)\,, \nonumber\\
\label{dPiNH} & & \\
N^{-m} Z^R_q  \bar{W}^R_{ q'} \Pi_N^\H(z,\theta;w,\phi)
&=& N^2 z_{q'}\bar w_q \Pi_N^\H(z,\theta;w,\phi) + N\de_{qq'}
\Pi_N^\H(z,\theta;w,\phi)\,.   \nonumber
\end{eqnarray} 

\begin{rem}  Recall that the
metric on $\ocal(N)\to\CP^m$ is given by $h^N(z)=(1+|z|^2)^{-N}$ using
the
coordinates and local frame from Example \ref{example1}. Since
$$h^N(z/\sqrtn)\to h_\H(z)\,,$$ the Heisenberg bundle can be regarded as
the
scaling limit of $\ocal(N)$.  (Of course, in the same way $L_\H$ is the
scaling limit of
$L^N$, for any positive line bundle $L\to M$.)

\end{rem}

\bigskip
\section{Correlation functions} \label{corfuns}

This section begins with a generalization to arbitrary dimension and
codimension a formula of \cite{H} and \cite{BD} for the ``correlation
density function'' in the one-dimensional case.  In fact, our formula
(Theorem~\ref{density}) applies to a general class of probability spaces
of
$k$-tuples of (real or complex) functions. 
We then specialize to the case where the space of sections
has a Gaussian measure. Finally, we show how the correlations of the
zeros of
$k$-tuples of sections of the $N$-th power of a holomorphic line bundle
are
given by a rational function in the Szeg\"o kernel $\Pi_N$ and its
derivatives (Theorem \ref{npointcor}).

\subsection {General formula for zero correlations}

For our general setting, we let
$(V,h)$ be a Hermitian holomorphic vector bundle on an
$m$-dimensional Hermitian complex manifold $(M,g)$. (Here, we make no
curvature assumptions.)  Suppose that
$\scal$ is a finite dimensional subspace of the space $H^0(M,V)$ of
global
holomorphic sections of $V$, and let $d\mu$ be a probability measure on
$\scal$ given by a semi-positive $\ccal^0$ volume form that
is
strictly positive in a neighborhood of $0\in\scal$. (We shall later
apply
our results to the case where
$V=L^N\oplus\cdots\oplus L^N$, for a holomorphic line bundle
$L$ over a compact complex manifold $M$, and $\scal=H^0(M,V)$  with a
Gaussian measure $d\mu$.  Our formulation involving general vector
bundles
allows us to reduce the study of $n$-point correlations to the case
$n=1$,
i.e., to expected densities of zeros.)

As in the introduction, we introduce the punctured product
$$M_n=\{(z^1,\dots,z^n)\in \underbrace{M\times\cdots\times
M}_n: z^p\ne z^q \ \ {\rm for} \ p\ne q\}\,,$$ and we write
$$s(z)=(s(z^1),\dots,s(z^n))\,,\ 
\nabla s(z)=(\nabla s(z^1),\dots,\nabla s(z^n))\,,\quad
z=(z^1,\dots,z^n)\in
M_n\,,$$  where $\nabla s(\zeta)\in T^*_\zeta\otimes V_\zeta$ is the
covariant derivative with respect to the Hermitian connection on $V$. We
define the map
$$\jcal:M_n \times\scal\to \big[(\C\oplus T^*_M)\otimes V\big]^n\,,\quad
\jcal (z,s)=(s(z),\nabla s(z))\,;$$ i.e., $\jcal (z,s)$ is the 1-jet of
$s$
at $z\in M_n$.

We write $g=\Re\sum g_{qq'}dz_q\otimes d\bar
z_{q'},h_{jj'}=h(e_j,e_{j,})$,
where $\{z_1,\dots,z_m \}$ are local coordinates in $M$
and $\{e_1,\dots,e_k \}$ is a local frame in
$V$ ($m=\dim M,\ k={\rm rank}\,V$). We let $G=\det (g_{qq'})$, $H=\det
(h_{jj'})$.  We let
$$d\zeta=\frac{1}{m!}\omega^m_{\zeta}=G(\zeta)\prod_{j=1}^m d\Re \zeta_j
d\Im
\zeta_j\,,\quad \zeta\in M$$ denote Riemannian volume in $M$, and we
write
\begin{equation}\label{dbda} x^p=\sum_j b^p_j e_j(z^p),\quad dx^p=H(z^p)
\prod_{j} d\Re b^p_j d\Im b^p_j \quad x^p\in V_{z^p}\,,\end{equation}
$$\xi^p=\sum_{j,q} a^p_{jq} dz_q\otimes e_j|_{z^p}, \quad d\xi^p=
G(z^p)^{-k}H(z^p)^{m} \prod_{j,q} d\Re a^p_{jq} d\Im
a^p_{jq}
\quad \xi_j\in (T^*_M\otimes V)_{z^p}\,.$$  The quantities $dx^p,\;
d\xi^p$
are the intrinsic volume measures on $V_{z^p}$ and $(T^*_M\otimes
V)_{z^p}$,
respectively, induced by the metrics $g,h$.

\begin{defin} Suppose that $\jcal $ is surjective.  We define the {\it
$n$-point
density\/}  $D_n(x,\xi,z)dxd\xi
dz$ of $\mu$ by
\begin{equation}\label{jointdist} \jcal _*(dz\times
d\mu)=D_n(x,\xi,z)dxd\xi
dz\,,\quad x=(x^1,\dots,x^n)\in
V_{z^1}\times\dots\times V_{z^n}\,,\end{equation} 
$$\xi=(\xi^1,\dots,\xi^n)\in
(T^*_M\otimes V)_{z^1}\times\dots\times(T^*_M\otimes V)_{z^n}\,,\
z=(z^1,\dots,z^n)\in M_n\,,$$
$$dx=dx^1\cdots dx^n\,,\quad d\xi=d\xi^1\cdots d\xi^n\,,\quad
dz=dz^1\cdots
dz^n\,.$$ In this case, for each
$z\in M_n$, the (vector-valued) random variable
$(s(z),\nabla s(z))$ has (joint) probability distribution
$D_n(x,\xi,z)dxd\xi$.\end{defin}

\begin{rem} If we let $n=1$ and fix a point $z\in M$, then the measure
$D(x,
\xi, z) dx d\xi$ is intrinsically defined as a measure on the space
$J^1_z(M,
V)$ of 1-jets of sections of $V$ at $z$.  Taking a section to its
$1$-jet at
$z$ defines a map $\jcal_z: \mathcal{S} \to J^1_z(M,V)$ and hence
induces a
measure $\jcal_{z*} \mu$ on $J^1_z(M,V)$ independently of any choices of
connections, coordinates or metrics.  Similarly for $n>1$, $D(x,
\xi, z) dx d\xi$ is an intrinsic measure on $\prod _{p=1}^n J^1_{z_p}(M,
V)$. \end{rem}

For a vector-valued 1-form $\xi\in  T^*_{M,z}\otimes V_{z}= {\rm
Hom}(T_{M,z},V_z)$, we let $\xi^*\in {\rm Hom}(V_{z},T_{M,z})$ denote
the adjoint to $\xi$ (i.e., $\langle\xi^*v,t \rangle=\langle v,\xi
t\rangle\,$), and we consider the
endomorphism
$\xi\xi^*\in {\rm Hom}(V_{z},V_{z})$.
In terms of local frames, if $$\xi=\sum_j \xi_j \otimes
e_j=\sum_{j,q}a_{jq}
dz_j\otimes e_j\,,$$ then
$$\xi^*=\sum_{j,q}\al_{jq}\frac{\d}{\d z_q}\otimes e_j^*\,,\qquad
\al_{jq}=\sum_{j',q'} h_{jj'}\ga_{q'q}\bar a_{j'q'}\,,$$
where $\big(\ga_{qq'}\big)=\big(g_{qq'}\big)^{-1}$; hence we have
\begin{equation}\label{endo} \xi\xi^*=\sum_{j,j',j'',q,q'} 
h_{j'j''}a_{jq}\ga_{q'q}\bar a_{j''q'}\, e_j\otimes e_{j'}^*
\,.\end{equation} Its determinant is given by
\begin{equation}\label{detendo}\det(\xi\xi^*)=H\det\left(\sum_{q,q'}
a_{jq}\ga_{q'q}\bar a_{j'q'}\right)_{1\le j,j'\le k}=H\det \langle\xi_j,
\xi_{j'}\rangle=
H\|\xi_1\wedge\dots\wedge \xi_k\|^2\,.\end{equation}

\begin{rem} The measure $\det(\xi \xi^*) D(0, \xi, z) d\xi dz$ will play
a
fundamental role in our study of correlation functions.  We observe here
that it depends only on the metric $\om$ on $M$, and in the case where
the
zero sets are points ($k=m$), it is
independent of the choice of metric on $M$ as well. 
Indeed, as mentioned in the previous remark,
$D(x,\xi, z) dx d\xi$ is well-defined on $J^1_z(M,V)$.
The conditional density $D(0, \xi, z) d\xi$ equals $\jcal_{z*} \mu /dx
|_{x =
0}$ and thus depends only on the choice of volume forms $dx^p$ on
$V_{z^p}$.
Since $dz/dx$ transforms in the opposite way to $\det \xi \xi^*$ it
follows
that  $\det(\xi \xi^*) D(0, \xi, z) d\xi dz$ is an invariantly defined
measure on $(T^*_M\otimes V)^n$. \end{rem}

Recall that for $s\in\scal$ so that $\codim Z_s=k$, we let $|Z_s|$
denote
Riemannian 
$(2m-2k)$-volume along the regular points of $Z_s$, regarded as a
measure on
$M$.

\begin{defin}  For $s\in\scal$ so that $\codim Z_s=k$, we consider
the product measure on $M_n$,
$$|Z_s|^n=\big(\underbrace{|Z_s|\times\cdots\times|Z_s|}_{n}\big)\,.$$
Its expectation $\E |Z_s|^n$ is called the {\it $n$-point zero
correlation measure.} \end{defin}

We shall use the following general formula to compute the correlations
of
zeros and to show universality of the scaling limit:

\begin{theo}\label{density} Let $M,V,\scal,d\mu$ be as above, and
suppose
that $\jcal $ is surjective and  the volumes $|Z_s|$ are locally
uniformly
bounded above.  Then
\begin{equation}\label{a8} \E|Z_s|^n =K_n(z)dz \,,\quad
K_n(z)=
\int d\xi\,D_n(0,\xi;z)\prod_{p=1}^n 
\det \left(\xi^{p}\xi^{p*}\right)\,.\end{equation}
\end{theo}

The function $K_{n}(z^1,\dots,z^n)$, which is continuous on $M_n$
is called the {\it $n$-point zero correlation function\/}.  For $k< m$,
(\ref{a8}) holds on all of the
$n$-fold product $M\times\cdots\times M$, including the diagonal, and
$K_n$ is
locally integrable on $M\times\cdots\times M$ (and is infinite on the
diagonal).  In the case
$k=m$, when the zero sets are discrete, the zero correlation measure on
$M\times\cdots\times M$ is the sum of the absolutely continuous measure
$K_n(z)dz$ plus a measure supported on the diagonal.

\medskip\noindent{\em Proof of Theorem \ref{density}:\/}
Consider the Hermitian vector bundle $V_n=\bigoplus_{p=1}^n
\pi_p^*V\longrightarrow M_n$, where $\pi_p:M_n\to M$ denotes the
projection
onto the $p$-th factor.  By replacing $V\to M$ with $V_n\to M_n$ and
$s\in H^0(M,V)$ with $$\tilde s(z^1,\dots,z^n)=(s(z^1),\dots,s(z^n))\in
H^0(M_n,V_n)\,,$$ and noting that $T_{M_n,z}=\prod_p T_{M,z^p}$ and
$|Z_s|^n=|Z_{\tilde s}|$, we can assume without loss of generality that
$n=1$.

It follows from the above remarks that $D(0,\xi;z)$ does not depend on
the
choice of connection on
$V$. We can also verify this in terms of local coordinates: write
$s=\sum
b_j e_j$,
$\nabla s =
\sum a_{jq} dz_q \otimes e_j$ as in (\ref{jointdist});
we have $a_{jq}=\frac{\d b_j}{\d z_q}
+ \sum_k b_k\theta^k_{jq}$. Then if we write $a^0_{jq}=\frac{\d b_j}{\d
z_q}$, we have $$\frac{\d(a_{jq},b_j)}{\d(a^0_{jq},b_j)} = 1\,.$$  Hence
$D(0,\xi;z)$ is unchanged if we substitute the (local) flat connection
given
by $a^0_{jq}$.

We now restrict to a coordinate neighborhood $U\subset M$
where $V$ has a local frame $\{e_j\}$.  By
hypothesis, we can suppose that the  $e_j$ are 
restrictions of sections in $\scal$.
We write
$s=\sum s_j e_j$, and by the above we may assume that $\nabla s =
\sum ds_j \otimes e_j$. 
We use the notation
$$\bbb\xi\bbb=\sqrt{\det(\xi\xi^*)}\,,
\quad {\rm for} \ \xi\in T^*_{M,z}\otimes V_{z}= {\rm
Hom}(T_{M,z},V_z)\,.$$
Then by (\ref{detendo}), $$\bbb\nabla s\bbb^2
=H\|
d s_1\wedge\cdots \wedge d s_k\|^2=\|\Psi\|\,,$$
where $\Psi$ is the $(k,k)$-form on
$U$ given by: $$\Psi=H
\left(\frac{i}{2}
ds_1\wedge
\overline{d s_1}\right)\wedge\cdots\wedge 
\left(\frac{i}{2} ds_k\wedge
\overline{d s_k}\right)\,.$$ Thus, by the Leray formula,
\begin{equation}\label{Leray}
|Z_s|=\|ds_1\wedge\cdots\wedge ds_k\|^2 \left.\frac{dz }{ 
\frac{i}{2}ds_1\wedge d\bar s_1\cdots\wedge\frac{i}{2}ds_k\wedge  d\bar
s_k}\right|_{Z_s}=\bbb\nabla s\bbb^2 \left.
\frac{dz}{\Psi}\right|_{Z_s}\,,\end{equation}

Define the measure $\la$ on $M\times\scal$ by
\begin{equation}\label{lambda} (\lambda,\phi)=\int_\scal
\left(|Z_s|,\phi(z,s)\right)d\mu(s)\,.\end{equation}  Then
$$\pi_*\la=\E|Z_s|^n\,,$$ where $\pi:M\times\scal\to M$ is the
projection.
Hence,
\begin{equation}\label{Leray2}\la=\int_\scal d\mu(s)\; |Z_s|
=\int_\scal d\mu(s)\left.\left(\bbb \nabla s\bbb^2\frac{dz}{
\Psi}\right)\right|_{Z_s}\,.
\end{equation}

For (almost all) $x\in\C^{k}$, let $I(s,x)$ be the measure on $U$ given
by
$$(I(s,x),\phi)=\int_{s(z)=\sum x_j e_j(z)}\phi(z) d\vol_{(2m-2k)n}(z)=
\int_{s(z)=\sum x_j e_j(z)}
\bbb \nabla s\bbb^2\frac{dz}{ \Psi}\phi(z)\,,\ \phi\in\ccal^0(U) \,,$$ 
where the second equality is by (\ref{Leray}) applied to $s-\sum x_j
e_j(z)$.
Then
\begin{equation}\label{I} \int
I(s,x) dx=\bbb \nabla s(z)\bbb^2 dz\,.\end{equation} 
Now let $\lambda_x$ be the measure on $U$ given by
$$(\lambda_x,\phi)=\int_\scal (I(s,x),\phi)d\mu(s) \,.$$

\begin{claim} The map
$x\mapsto (\la_x,\phi)$ is continuous.\end{claim}

\noindent To prove this claim, we first note that the hypothesis that
$|Z_s|$
is locally uniformly bounded implies that $(I(s,x),\phi)\le C<+\infty$
uniformly in $s,x$.  Thus we can assume without loss of generality that
$\mu$ has compact support in $\scal$.  By hypothesis, the map
$$\sigma:U\times
\scal\to
\C^k,\qquad \sigma(z,s)= (s_1(z),\dots,s_k(z))$$ is a submersion. We can
now write $\la_x$ as a fiber integral of a compactly supported $\ccal^0$
form:
$$\la_x=\frac{1}{(m-k)!}\int_{\sigma^{-1}(x)}\phi(z)\om^{m-k}(z)\wedge
d\mu(s)\,,$$ and thus $\la_x$ is continuous, verifying the claim.

\medskip We note that
$\lambda_0=\la|_U$. Hence, to complete the proof, we must show that
$$\pi_* \la_0=K_1(z)dz|_U\,.$$

By (\ref{jointdist}) and (\ref{I}), for a test function $\phi(x,\xi,z)$,
\begin{eqnarray*}
\int
\phi(x,\xi,z)\bbb\xi\bbb^2 D_1(x,\xi,z)dxd\xi dz &=& \int
d\mu(s)\int \phi(\jcal (z,s))
\bbb \nabla s(z)\bbb^2dz\\ &=&\int dx\int 
(I(s,x),{\phi\circ \jcal }) d\mu(s)\\ &=&\int
(\la_x,\phi\circ \jcal ) dx\,.\end{eqnarray*} 
By choosing $\phi(x,\xi,z)=\rho_\ep(x)\psi(z)$, where $\rho_\ep$ is an
approximate identity, and letting $\ep\to 0$, we conclude that $$\int 
\psi(z)K_1(z)dz=\int
\psi(z)\bbb\xi\bbb^2 D_1(0,\xi,z)d\xi dz
=(\la_0,\psi(z))\,.$$
\qed\medskip

We note the following analogous formula for real
manifolds:

\begin{theo}\label{density-real} Let $V$ be a $\ccal^\infty$ real vector
bundle over a
$\ccal^\infty$  Riemannian manifold $M$, and let $\mu$ be
a probability measure on a  finite dimensional vector space
$\scal$ of $\ccal^\infty$ sections of $V$ given by a semi-positive
volume
form that is strictly positive at $0$.  Suppose that the volumes $|Z_s|$
are
locally uniformly bounded above.  Let $D_n(x,\xi,z)dxd\xi dz$ denote the
$n$-point density of $\mu$.  Then
\begin{equation}\label{a8-real} \E|Z_s|^n =K_n(z)dz \,,\quad
K_n(z)=
\int d\xi\,D_k(0,\xi,z)\prod_{p=1}^n 
\sqrt{\det(\xi^{p}\xi^{p*})}\,.
\end{equation} \end{theo}

The proof is similar to that of Theorem \ref{density}, except that
(\ref{Leray}) is replaced by the Leray formula 
\begin{equation}\label{Leray-real} |Z_s|=\|ds_1\wedge\cdots\wedge ds_k\|
\left.\frac{d\zeta }{  ds_1\wedge\cdots\wedge
ds_k}\right|_{Z_s}\end{equation} in the real case.

\subsection{Formula for Gaussian densities}\label{formula-gaussian} We
now
specialize our formula from Theorem
\ref{density} to the case where $\mu$ is a Gaussian measure. Fix
$z=(z^1,\dots,z^n)\in M_n$ and choose local coordinates $\{z^p_q\}$ and
local frames  $\{e^p_j\}$ near $z^p$, $p=1,\dots,n$. We consider the
random
variables
$b^p_j,\ a^p_{jq}$ given by 
\begin{equation}\label{fg1} s(z^p)=\sum_{j=1}^k b^p_j e^p_j,\quad
\nabla s(z^p)=\sum_{j=1}^k\sum_{q=1}^m a^p_{jq}dz^p_q\otimes
e^p_j,\qquad
 p=1,\dots,n.\end{equation}  By
(\ref{gaussian})--(\ref{gaussian2}) and (\ref{dbda})--(\ref{jointdist})
the
$n$-point density
$$D_n(x,\xi,z)dxd\xi
dz=D_n\left[\prod_{p=1}^nG(z^p)^{-k}H(z^p)^{m}\right]
dbdadz$$  is given by:
\begin{equation}\label{fg2}D_{n}(b,a;z)=
\frac{\exp\langle-\De_{n}^{-1}v,
v\rangle}{\pi^{kn(1+m)}\det\De_{n}}\;,\qquad v=\pmatrix b\\
a\endpmatrix\,, 
\end{equation}
where \begin{equation}\label{fg3}
\Delta_{n}=\left(
\begin{array}{cc}
A_{n} & B_{n} \\
B^*_{n} & C_{n}
\end{array}\right)\end{equation}
$$A_{n}=\big(A^{jp}_{j'p'}\big)=
\big (\,\E b^p_j \bar b^{p'}_{j'}\,\big),\quad
B_{n}=\big(B^{jp}_{j'p'q'}\big)=
\big (\,\E  b^p_j \bar a^{p'}_{j'q'}\,\big),\quad
C_{n}=\big(C^{jpq}_{j'p'q'}\big)=
\big(\,\E   a^{p}_{jq}\bar a^{p'}_{j'q'}\,\big);$$
$$j,j'=1,\dots,k;\quad p,p'=1,\dots,n; \quad q,q'=1,\dots,m.
$$
(We note that $A_n,\ B_n,\ C_n$ are $kn\times kn,\ kn\times knm,\ 
knm\times knm$ matrices, respectively; $j,p,q$ index the rows, and 
$j',p',q'$ index the columns.)

The function $D_{n}(0,a;z)$ is a Gaussian function, but it is not
normalized as a probability density. It can be represented as
\begin{equation}\label{fg4}
D_{n}(0,a;z)=Z_{n}(z) D_{\La_n}(a;z),
\end{equation}
where
\begin{equation}\label{fg5}
D_{\La_n}(a;z)=\frac{1}{\pi^{knm}\det\La_{n}}\exp\left(
-{\langle\La^{-1}_{n}a, a\rangle}\right)
\end{equation}
is the Gaussian density with covariance matrix
\begin{equation}\label{fg6}
\La_{n}=C_{n}-B^*_{n}A^{-1}_{n}B_{n} =\left(C^{jpq}_{j'p'q'}
-\sum_{j_1,p_1,j_2,p_2}\bar 
B_{jpq}^{j_1p_1}\Gamma^{j_1p_1}_{j_2p_2}B^{j_2p_2}_{j'p'q'}\right)
\qquad (\Gamma=A_n^{-1})
\end{equation}
and 
\begin{equation}\label{fg7}
Z_{n}(z)=\frac{\det\Lambda_{n}}{\pi^{kn}\det\De_{n}}
=\frac{1}{\pi^{kn}\det  A_{n}}\,. 
\end{equation}

This reduces formula (\ref{a8}) to
\begin{equation}\label{fg8}
K_{n}(z)=\frac{1}{\pi^{kn}\det A_{n}}\left\langle
\prod_{p=1}^n \det\left(a^{p*}\gamma^pa^p\right)\right\rangle_{\La_{n}} 
\end{equation}
where $\langle\cdot\rangle_{\La_{n}}$ stands for averaging with
respect to the 
Gaussian density $D_{\La_n}(a;z)$, and $(\gamma^p_{qq'})=
(g^p_{qq'})^{-1}$,
$g^p_{qq'}=g_{qq'}(z^p)$. 

\subsection{Densities and the Szeg\"o kernel}\label{D-meet-S} We return
to
our positive Hermitian line bundle $(L,h)$ on a compact complex manifold
$M$
with \kahler form $\omega=\frac{i}{2}\Theta_h$. We now apply formulas
(\ref{fg3})--(\ref{fg8}) to the vector bundle
$$V=\underbrace{L^N\oplus\cdots\oplus L^N}_k$$
and space of sections  $$\scal=H^0(M,V)=H^0(M,L^N)^k$$ with the Gaussian
measure $\mu=\nu_N\times\cdots\times\nu_N$, where $\nu_N$ is the
standard
Gaussian
measure on $H^0(M,L^N)$ given by (\ref{gaussian}). We denote the
resulting
$n$-point density by $D_{nk}^N$, and we also write $\De_n=\De_{nk}^N,\
A_n
=A_{nk}^N$, etc.

As above, we fix
$z=(z^1,\dots,z^n)\in M_n$ and choose local coordinates $\{z^p_q\}$ near
$z^p$, $p=1,\dots,n$. We also choose local frames  $\{e^p_L\}$ for $L$
near
the points $z^p$ so that
$$\|e^p_L(z^p)\|_h=1\,.$$ For
$s\in\scal$, we write
\begin{equation}\label{dms1} s(z^p)=
\left(
\begin{array}{c}
s_1(z^p)\\
\vdots\\
s_k(z^p)
\end{array}
\right)=
\left(
\begin{array}{c}
b^p_1\\
\vdots\\
b^p_k
\end{array}
\right) (e^p_L(z^p))^{\otimes N}
\,,\end{equation}
\begin{equation}\label{dms2} \nabla_N s_j(z^p)
=\sum_{q=1}^m a^p_{jq}dz^p_q\otimes (e^p_L(z^p))^{\otimes
N}\,.\end{equation}
Since the
$s_j$ are independent and have identical distributions, we have
\begin{equation}\label{dms3} A_{nk}^N=\big(A^{jp}_{j'p'}\big)=
\big(\de_{jj'}\E(b^p_1 \bar
b^{p'}_1)\big)\,,\quad B_{nk}^N=\big(B^{jp}_{j'p'q'}\big)=
\big(\de_{jj'}\E(b^p_1 \bar
a^{p'}_{1q'})\big)\,,
\end{equation}
$$C_{nk}^N=\big(C^{jpq}_{j'p'q'}\big)=
\big(\de_{jj'}\E(a^p_{1q} \bar
a^{p'}_{1q'})\big)\,.$$
We write $$s_1=\sum_{\al=1}^{d_N}c_\al
S^N_\al=\left(\sum_{\al=1}^{d_N}c_\al f^p_\al\right)(e^p_L)^{\otimes
N}\,,$$ where
$\{S^N_\al\}$ is an orthonormal basis for $H^0(M,L^N)$. 
Using the local coordinates $(z^p,\theta)$ in $X$ as described in \S
\ref{cxgeom}, we have by (\ref{dms3}) and (\ref{PiN}) (noting that
$h(z^p)=0$ by the above choice of local frames),
\begin{equation}\label{dms4}A^{jp}_{j'p'}
=\de_{jj'}\sum_{\al,\be=1}^{d_N}\E(c_\al \bar
c_\be)f^p_\al(z^p) \overline{f^{p'}_\be(z^{p'})}=\de_{jj'}
\sum_{\al=1}^{d_N}f^p_\al(z^p) \overline{f^{p'}_\al(z^{p'})}=\de_{jj'}
\Pi_N(z^p,0;z^{p'},0)\,.\end{equation}
Similarly, \begin{equation}\label{dms5}
B^{jp}_{j'p'q'}=\de_{jj'}
\sum_{\al=1}^{d_N}f^p_\al(z^p)
\overline{f^{p'}_{\al;q'}(z^{p'})}=\de_{jj'}
d^H_{\bar w_{q'}}\Pi_N(z^p,0;z^{p'},0)\,,\end{equation}
\begin{equation}\label{dms6}
C^{jpq}_{j'p'q'}=\de_{jj'}
\sum_{\al=1}^{d_N}f^p_{\al;q}(z^p)
\overline{f^{p'}_{\al;q'}(z^{p'})}=\de_{jj'}
d^H_{z_{q}}d^H_{\bar w_{q'}}\Pi_N(z^p,0;z^{p'},0)\,.\end{equation}

\begin{lem}\label{detN} There is a positive integer $N_0=N_0(M,n)$ such
that 
$$\det\big(\Pi_N(z^p,0;z^{p'},0)\big)_{1\le p,p'\le n} \ne 0\,,$$
for distinct points $z^1,\dots,z^n$ of $M$ and for all $N\ge N_0$.
\end{lem}

\begin{proof} It is a well-known consequence of the Kodaira Vanishing
Theorem
(see for example,
\cite{GH}) that we can find $N_0$ such that if $N\ge N_0$ and 
$x_1,\dots,x_n\in M$ with $x_p\ne x_1$ for $2\le p\le n$, then there is
a
section $s\in H^0(M,L^N)$ with $s(x_1)\ne 0$ and $s(x_p)=0$ for $2\le
p\le
n$.

We write $\wt A_{pp'}=\Pi_N(z^p,0;z^{p'},0)$. Suppose on the
contrary that
$\det (\wt A_{pp'})=0$, and chose a nonzero vector
$v=(v_1,\dots,v_n)$ such that $\sum_p v_p\wt A_{pp'}=0$.
Then recalling (\ref{szego}), we have
\begin{equation}\label{det1} 0=\sum_{p,p'} v_p \wt A_{pp'}\bar v_{p'}
= \sum_{p,p',\al} v_p \wh S^N_\al(z^p,0) \overline{\wh
S^{N}_\al(z^{p'},0)
v_{p'}} =\sum_{\al=1}^{d_N} |x_\al|^2\,,\end{equation}
where $x_\al=\sum_p v_p\wh S^N_\al(z^p,0)$. Since the $S^N_\al$ span
$H^0(M,L^N)$, it follows that for all $s\in H^0(M,L^N)$, we have $\sum_p
v_p \hat s(z^p) =0$.
But this contradicts the fact that, choosing $p_0$ with $v_{p_0}\ne
0$, we can find a section $s\in H^0(M,L^N)$ with
$s(z^{p_0})\ne 0$ and $s(z^p)=0$ for $p\ne p_0$.
\end{proof}

Thus we see that the $n$-point correlation functions depend only on the
Szeg\"o kernel, as follows:

\begin{theo}\label{npointcor} Let $(L,h)$ be a positive Hermitian line
bundle
on an
$m$-dimensional compact complex manifold $M$ with \kahler form
$\om=\frac{i}{2}\Theta_h$, let
$\scal=H^0(M,L^N)^k$ ($k\ge 1$), and give $\scal$ the standard Gaussian
measure
$\mu$ described above. Let $n\ge 1$ and suppose that $N$ is sufficiently
large so that 
$\jcal$ is surjective. Let $z=(z^1,\dots,z^n)\in M_n$ and choose local
coordinates $(\zeta_1,\dots,\zeta_m)$ at each point $z^p$ such that
$\Theta_h(z^p)=\sum_q d\zeta_q\wedge d\bar\zeta_q$, $1\le p\le n$.
Then the $n$-point correlation $K_{nk}^N(z)$ is given by a universal
rational
function, homogeneous of degree $0$, in the values of $\Pi_N$ and its
first
and second derivatives at the points $(z^p,z^{p'})$. Specifically,
\begin{equation}\label{npointcoreq}
K_{nk}^N(z)= \frac{\pcal_{nkm}\big(\Pi_N(z^p,z^{p'}),d^H_{\bar
w_{q}}\Pi_N(z^p,z^{p'}), d^H_{z_{q}}\Pi_N(z^p,z^{p'}),
d^H_{z_{q}}d^H_{\bar
w_{q'}}\Pi_N(z^p,z^{p'})\big)}{\pi^{kn}\left[\det
\big(\Pi_N(z^p,z^{p'})\big)_{1\le p,p'\le
n}\right]^{k(n+1)}}\end{equation}
($1\le p,p'\le
n,\ 1\le q,q'\le m$), where $\pcal_{nkm}$ is a universal homogeneous
polynomial of degree $kn(n+1)$ with integer
coefficients depending only on $n,k,m$.
\end{theo}

\begin{proof} The $n$-point zero correlation $K_{nk}^N(z)$ is given by
equation (\ref{fg8}) with
$\gamma^p_{qq'}=\de_{qq'}$. 
By the Wick formula (\cite[(I.13)]{Si}), the expectation
$$\left\langle
\prod_{p=1}^n \det\left(a^{p*}a^p\right)\right\rangle_{\La_{n}}$$ in
(\ref{fg8}) is a homogeneous polynomial (over $\Z$) of degree $kn$ in
the
coefficients of 
$\La_{n}$. By  (\ref{fg6}) and (\ref{dms4}), the coefficients of $\det
\big(\Pi_N(z^p,z^{p'})\big)\La_n$ are homogeneous polynomials of degree
$n+1$
in the coefficients of $A_n,B_n,C_n$.  The conclusion then follows from
(\ref{dms4})--(\ref{dms6}). 
\end{proof}

\begin{rem} 
In the statement of Theorem \ref{npointcor}, we wrote $\Pi_N(z,w)$ for
$\Pi_N(z,\theta;w,\phi)$.  Since the expression is homogeneous of degree
0, it
is independent of $\theta$ and $\phi$.  Alternately, we could regard
$\Pi_N(z,w)$ as functions on $M\times M$ having values in
$L_z\otimes\overline{L_w}$ (replacing the horizontal derivatives with
the
corresponding covariant derivatives); again the degree 0 homogeneity
makes the
expression a scalar.  Furthermore, since Theorem \ref{density} is valid
for
all connections, we can replace the horizontal derivatives in
(\ref{npointcoreq}) with the derivatives with respect to an arbitrary
connection.\end{rem}

\subsection {Zero correlation for 
$\SU(m+1)$-polynomials} \label{zcorpoly}

In this section, we use our methods to describe the zero correlation
functions for $\SU(m+1)$-polynomials.  We do not carry out the
computations in
complete detail, since we are primarily interested in the scaling
limits,
which we shall compute in \S \ref{formulas}.

The $\SU(m+1)$-polynomials are random homogeneous
polynomials of degree $N>0$ on $\C^{m+1}$,
\begin{equation}\label{a1}
s(z)=s(z_0,z_1,\dots,z_m)=\sum_{|J|=N}
\sqrt{N!/J!}\,{c_J}z^J,\quad z^J=z_0^{j_0}\cdots
z_m^{j_m},\quad  J!=j_0!\cdots j_m!,
\end{equation}
where the coefficients $c_J$ are complex independent Gaussian random
variables with mean 0 and variance 1:
\begin{equation}\label{a2}
\E  c_J=0;\qquad \E  c_J\overline{c_K}=
\de_{JK},\quad \de_{JK}=\de_{j_0k_0}\dots\de_{j_mk_m};
\qquad \E  c_Jc_K=0.
\end{equation}
Then $s(z)$ is a Gaussian random polynomial on
$\C^{N+1}$ with first and second moments given by
\begin{equation}\label{a3}
\E  s(z)=0;\qquad \E  s(z)\overline{s(w)}=\langle z, w
\rangle^N =\left(\sum_{q=0}^m z_q\overline{w_q}\right)^N;
\qquad \E  s(z)s(w)=0.
\end{equation}
This implies that
the probability distribution of $s(z)$ is invariant with respect
to the map $s(z)\to s(Uz)$ for all $U\in \SU(m+1)$.

Let $(\scal_N,\mu_N)$ denote the Gaussian probability space of
independent $k$-tuples ($k\le m$) of
$\SU(m+1)$-polynomials of degree $N$.  For 
$s=(s_1,\dots,s_k)\in \scal_N$, the zero set
\[
Z_s=\{z: s_1(z)=\dots=s_k(z)=0\}\,.
\] 
is an algebraic variety in the complex projective space $\CP^m$.
We will assume that $\CP^m$ is supplied with the Fubini-Study
Hermitian metric $\om$, which is $\SU(m+1)$-invariant. In the affine
coordinates $z=(1,z_1,\dots,z_m)$,
\begin{equation}\label{a3a}
\om=\frac{\sqrt{-1}}{ 2}\ddbar\log\left(1+\sum|z_q|^2\right)=\frac
{\sqrt{-1}}{
2}\left[\frac{\sum dz_q\wedge\overline{dz_q}}{1+\sum |z_q|^2}
-\frac{\left(\sum \overline{z_q}dz_q\right)\wedge \left(\sum
z_q\overline{dz_q}\right)} {\left(1+\sum |z_q|^2\right)^2}\right]\,;
\end{equation}
i.e.,
\begin{equation}\label{FSg} \om=
\frac{\sqrt{-1}}{ 2}\sum g_{qq'}dz_q\wedge
dz_{q'}\,, \quad 
g_{qq'}=\frac{(1+|z|^2)\de_{qq'}-\bar
z_qz_{q'}}{(1+|z|^2)^2}\,.\end{equation}

To simplify our computations, we consider only points $z^p$ with finite
affine
coordinates, $z^p=(1,z^p_1,\dots,z^p_m),\; p=1,\dots,n$, and we regard
the
$\SU(m+1)$-polynomials $s_j$ as polynomials of degree $\le N$ on $\C^m$;
i.e.,
we regard the $s_j$ as sections of the trivial line bundle on $\C^m$
with
the flat metric $h=1$ (so that the covariant derivatives coincide with
the
usual derivatives of functions).

As above, we consider the
random variables $$b^p_j=s_j(z^p),\; a^p_{jq}=\frac{\d s_j }{ \d
z_q}(z^p)\,,$$
and we denote their joint
distribution by
\begin{equation}\label{a11a}
\begin{array}{rl}
D_{nk}^N(b,a;z)db\,da,\quad  & b=\left(b^1,\dots,b^n\right),\quad
b^p=
(b^p_j)_{j=1,\dots,k};\\
&\\
&  a=\left(a^1,\dots,a^n\right),\quad
a^p=\left(a^p_{jq}\right)_{j=1,\dots,k;\;q=1,\dots,m}\,.
\end{array} 
\end{equation}
(Here, the $n$-point density is with respect to Lebesgue measure
$db=\prod
db^p_j d\bar b^p_j,\; da=\prod da^p_{jq}d\bar a^p_{jq}$.) We assume that
$N>nm$ to ensure that $\mu_N$
possesses a continuous $n$-point density.  
Since $\mu_N$ is Gaussian, the density $D_{nk}^N(b,a;z)$ is Gaussian as
well,
and it is described by the covariance matrix
\begin{equation}\label{a14}
\Delta_{nk}^N=\left(
\begin{array}{cc}
A_{nk}^N & B_{nk}^N \\
B_{nk}^{N*} & C_{nk}^N
\end{array}\right)
\end{equation}
where
\begin{equation}\label{a14a}
\begin{array}{l}
A_{nk}^N=\left (\,\E  s_j(z^p)\overline{s_{j'}(z^{p'})}\,\right),\\
B_{nk}^N=\left (\,\E  s_j(z^p)\overline{\frac{\d s_{j'}}{\d
    z_{q'}}(z^{p'})}\,\right),\\ 
C_{nk}^N=\left (\,\E  \frac{\d s_j}{\d z_q}(z^p)
\overline{\frac{\d s_{j'}}{\d
    z_{q'}}(z^{p'}  )}\,\right);\\
j,j'=1,\dots,k;\quad p,p'=1,\dots,n; \quad q,q'=1,\dots,m.
\end{array}
\end{equation}

By (\ref{a14a}) and (\ref{a3}),
\begin{equation}\label{a23}
\begin{array}{ll}
A_{nk}^N=\left(\de_{jj'}S_N(z^p,z^{p'})\right)\,,&\displaystyle
S_N(z,w)=\left(1+\sum_{r=1}^m z_r\overline{w_r}\right)^N,\\
B_{nk}^N=\left(\de_{jj'}S_{Nq'}(z^p,z^{p'})\right)\,,
&\displaystyle
S_{Nq'}(z,w)=Nz_{q'}\left(1+\sum_{r=1}^m
z_r\overline{w_r}\right)^{N-1}\,, \\ 
C_{nk}^N=\left(\de_{jj'} S_{Nqq'}(z^p,z^{p'})\right)\,,&\displaystyle
S_{Nqq'}(z,w)=N(N-1)\overline{w_q}z_{q'}
\left(1+\sum_{r=1}^m z_r\overline{w_r}\right)^{N-2}\\ 
&\displaystyle \qquad +\de_{qq'}N\left(1+\sum_{r=1}^m
z_r\overline{w_r}\right)^{N-1}
\end{array}
\end{equation}

The $n$-point zero correlation functions $K_{nk}^N$ for the
$\SU(m+1)$-polynomial $k$-tuples $\scal_{N}$ can be computed by
substituting
(\ref{a23}) into formulas (\ref{fg6}) and (\ref{fg8}).  (Alternately, we
can
compute the zero correlation functions with respect to the Euclidean
volume
on $\C^m$ by setting $\gamma^p=$Id in (\ref{fg8}).)

\begin{rem} Note that the one-point correlation
function, or the zero-density function, is constant, since it is
invariant
with respect to the group $\SU(m+1)$.  Indeed,
by B\'ezout's theorem and
(\ref{volmeasure}),
\begin{equation}\label{a4}
|Z_s|(1)=\vol(V_s)=\int_{V_s}{\textstyle\frac{1}{ (m-k)!}}\om^{m-k}
=\Omega_{2m-2k}\deg V_s=\Omega_{2m-2k}N^k\;,
\end{equation}
where \begin{equation}\label{a4a}\Omega_{2\ell} = \vol\, \CP^\ell =
\frac{\pi^\ell}{
\ell !}\,.\end{equation} Hence,
\begin{equation}\label{a32b}
K_{1k}^N(z)=\frac{\vol\, Z_s}{\vol\,
\CP^m}=\frac{N^{k}\Omega_{2m-2k}}{\Omega_{2m}}
=\frac{N^{k}m!}{(m-k)!\pi^k}\,.\end{equation} We can also use our
formulas to
compute $K_{1k}^N$ directly:  By (\ref{a23}),
\begin{equation}\label{a32c}
A_{1k}^N=\Big(\de_{jj'}(1+|z|^2)^N\Big)\,,\quad
B_{1k}^N=\Big(\de_{jj'}Nz_{q'}(1+|z|^2)^{N-1}\Big)\,,\end{equation}
$$C_{1k}^N=\Big(\de_{jj'}N[(N-1)\bar
z_qz_{q'}+(1+|z|^2)\de_{qq'}](1+|z|^2)^{N-2}\Big)\,.$$ Hence by
(\ref{fg6}),
\begin{equation}\label{a32d} \La_{1k}^N =
\Big(\de_{jj'}N[(1+|z|^2)\de_{qq'}-\bar
z_qz_{q'}](1+|z|^2)^{N-2}\Big)=
\Big(\de_{jj'}N(1+|z|^2)^Ng_{qq'}(z)\Big)\,.\end{equation}
In the hypersurface case ($k=1$), we compute
$$K_{11}^N=\frac{1}{ \pi (1+|z|^2)^N}\left.\left\langle\sum_{q,q'=1}^m
\bar a_{1q}\gamma_{qq'} a_{1q'}\right\rangle\right|_{\La_{11}^N}=
\frac{N}{
\pi}\sum_{q,q'=1}^m \gamma_{qq'}g_{q'q}=\frac{Nm}{ \pi}\,,$$
as expected. For $k>1$, we have
$\La_{1k}^N(0)=NI$ where
$I$ is the unit matrix, and (\ref{fg8}) yields
$$
K_{1k}^N(0)=\frac{N^{k}}{\pi^k}\left\langle
\det\left(\sum_{q=1}^m
\bar a_{jq}{a_{j'q}}\right)_{j,j'=1,\dots,k} \right\rangle_I
=\frac{N^{k}m!}{(m-k)!\pi^k}\,,$$
which agrees with (\ref{a32b}).
\end{rem}

\bigskip
\section{Universality and scaling}

Our goal is to derive scaling limits of the $n$-point correlations
between the
zeros of {\it random} $k$-tuples of sections of powers of a positive
line
bundle over a complex manifold. We expect the scaling limits to exist
and to
be universal in the sense that they should depend only on the dimensions
of
the algebraic variety of zeros and the manifold. Our plan is the
following. We first describe scaling in the Heisenberg model, which we
use
to provide the universal scaling limit for the
Szeg\"o kernel (Theorem \ref{neardiag}). Together with Theorem
\ref{npointcor}, this demonstrates the universality of the scaling-limit
zero
correlation in the case of powers of any positive line bundle on any
complex
manifold.

\subsection{Scaling of the Szeg\"o kernel in the Heisenberg group}
Our model for scaling is the Szeg\"o kernel for the reduced Heisenberg
group
described in
\S \ref{example2}.  Recall that for the simply-connected Heisenberg
group
${\bf H}^m$,  the scaling operators 
(or Heisenberg dilations)
$$ \delta_r (\zeta, t) = (r \zeta, r^2 t)\,,\quad r
\in \R^+$$
are automorphisms of ${\bf H}^m$ (\cite{F} \cite{Stein}). Since the
Szeg\"o kernel $\Pi$ of ${\bf H}^m$ is the unique self-adjoint
holomorphic
reproducing kernel, it follows that it must be invariant (up to a
multiple) under these automorphisms. In fact, one has
(\cite[p.~538]{Stein}):
\begin{equation} \label{SI}
\Pi(\de_rx,\de_ry)=r^{-2m-2}\Pi(x,y)\end{equation}

The condition for a dilation $\delta_r$ to descend to the quotient
group ${\bf H}_{\rm red}^m$ is that
$r^2 \Z \subset \Z$, or equivalently,
$r= \sqrt{N}$
with $N \in \Z^+$. Note however that $\de_{\sqrtn}$ is not an
automorphism of ${\bf H}^m_{\rm red}$ and there
is no well-defined dilation by $\sqrt{N}^{-1}$. 

The scaling identity (\ref{SI}) descends to ${\bf H}^m_{\rm red}$ in
the form \begin{equation}\Pi^\H_N(x, y) = N^m \Pi^\H_1 (\de_{\sqrtn}\,
x,
\de_{\sqrtn}\, y)\end{equation}
with \begin{equation}\label{sksl} \Pi^\H_1(x, y) =\frac{1}{\pi^m}
e^{i(\theta - \phi)}e^{ i \Im (z \cdot \bar{w})} e^{- 
\half |z - w|^2}\,,\quad x=(z,\theta)\,,\ y=(w,\phi)\,. \end{equation} 
(Recall (\ref{PiHN}).) Informally, we may say that the scaling limit of
$\Pi^\H_N$
equals $\Pi^\H_1$. Since scaling by $\sqrt{N}^{-1}$ is not well-defined
on ${\bf H}_{\rm red}^m$ it is more correct to say that $\Pi^\H_N$ is
the
$\sqrt{N}$ scaling of the scaling limit kernel.

\subsection{Scaling limit of a general Szeg\"o kernel}
We now show that $\Pi^\H_1$ is the scaling limit of the
$N$-th Szeg\"o kernel $\Pi_N$ of an arbitrary positive line bundle $L
\to
M$ in
the sense of the following ``near-diagonal asymptotic estimate for the
Szeg\"o kernel."
 
\begin{theo} \label{neardiag} 
Let $z_0\in M$ and choose local coordinates in a neighborhood of
$z_0$ so that $\Theta_h(z_0)=\sum dz_j\wedge d\bar z_j$.  Then
\begin{eqnarray*}N^{-m}\Pi_N(z_0+\frac{u}{\sqrtn},\frac{\theta}{N};
z_0+\frac{v}{\sqrtn},\frac{\phi}{N})&=&
\frac{1}{\pi^m} e^{i(\theta-\phi)+i\Im
(u\cdot \bar v)-\half |u-v|^2} + O(N^{-1/2})\\ &=&
\Pi^\H_1(u,\theta;v,\phi) + O(N^{-1/2})\;.\end{eqnarray*}
\end{theo}

To prove Theorem \ref{neardiag}, we need to recall the Boutet de
Monvel-Sjostrand parametrix construction:

\begin{theo} {\rm \cite [Th.~1.5 and \S 2.c]{B.S}}   Let $\Pi(x,y)$ be
the
Szeg\"o kernel of the boundary $X$ of a bounded strictly pseudo-convex
domain
$\Omega$ in a complex manifold $L$.   Then: there exists a symbol $s \in
S^{n}(X \times X \times \R^+)$ of the type $$s(x,y,t) \sim
\sum_{k=0}^{\infty} t^{n-k} s_k(x,y)$$ so that $$\Pi (x,y) =
\int_0^{\infty}
e^{i t\psi(x,y)} s(x,y,t ) dt $$ where the phase $\psi \in C^{\infty}(X
\times X)$ is determined  by the following properties:

$\bullet$ $\psi(x,x) = \frac{1}{i} \rho(x)$ where $\rho$ is the defining
function of $X$.

$\bullet$ $\bar{\partial}_x \psi$ and $\partial_y \psi$ vanish to
infinite
order along the diagonal.

$\bullet$ $\psi(x,y) = -\bar{\psi}(y,x)$. \end{theo}

The integral is defined as a complex oscillatory integral and is
regularized
by taking the principal value (see \cite{B.S}). The phase is determined
only
up  to a function which vanishes to infinite order at $x = y$ and its
Taylor
expansion  at the diagonal is given by \begin{equation} \psi(x + u, x +
v) =
\frac{1}{i} \sum \frac{\partial^{\alpha + \beta} \rho} {\partial
z^{\alpha}\partial \bar{z}^{\beta}}(x)
\frac{u^{\alpha}}{\alpha!}\frac{\bar{v}^{\beta}}{\beta!}.\end{equation}

The Szeg\"o kernels $\Pi_N$ are Fourier coefficients of $\Pi$ and hence
may be
expressed as: \begin{equation}\Pi_N(x,y) = \int_0^{\infty} \int_0^{2\pi}
e^{- i
N \theta}  e^{it  \psi( r_{\theta} x,y)} s(r_{\theta} x,y,t)
d\theta dt \end{equation} where $r_{\theta}$ denotes the $S^1$ action
on $X$. Changing variables $t \mapsto N t$ gives
\begin{equation} \Pi_N(x,y) = N \int_0^{\infty} \int_0^{2\pi}
e^{ i N ( -\theta + t \psi( r_{\theta} x,y))} s(r_{\theta} x,y, t
N) d\theta dt\,.\end{equation}

We now fix $z_0$ and consider the asymptotics of 
 \begin{equation}\begin{array}{l}\displaystyle \Pi_N(z_0 +
\frac{u}{\sqrt{N}}, 0; z_0 + \frac{v}{\sqrt{N}}, 0)
\\[14pt]\displaystyle
\quad\quad=
 N \int_0^{\infty} \int_0^{2\pi}
 e^{ i N ( -\theta + t\psi( z_0 + \frac{u}{\sqrt{N}}, \theta; z_0 +
\frac{v}{\sqrt{N}}, 0))} s(z_0 + \frac{u}{\sqrt{N}}, \theta; z_0 +
\frac{v}{\sqrt{N}}, 0), t N) d\theta dt \,.\end{array}\end{equation}

In our setting the phase takes the following concrete form: We
let $h (z,\bar{w})$ be the almost analytic function on $M \times M$
satisfying
$h(z,\bar{z}) = \|e_L\|^{-2}_h(z)$.  The
function $h(z,\bar{w})$ is defined by \begin{equation}\label{taylorh}
h (z_0 + u, \bar z_0 +
\bar v) =  \sum \frac{\partial^{\alpha + \beta} h(z,\bar{z})} {\partial
z^{\alpha} \partial \bar{z}^{\beta}} (z_0) \frac{u^{\alpha}}{\alpha!} 
\frac{\bar{v}^{\beta}}{\beta!}. \end{equation}

We consider the complex manifold $Y=L^*$ 
and we let $(z,\lambda)$ denote the
coordinates of $\xi \in Y$ given by $\xi = \lambda (e_L^*)_z$. In the
associated coordinates  $(x,y) = (z, \lambda, w, \mu)$ on $Y \times Y$,
we
have: \begin{equation} \rho(z,\lambda) =  1 - h(z,
\bar{z})|\lambda|^2,\;\;\;\; \psi(z, \lambda, w, \mu) = \frac{1}{i}(1 - 
h(z,\bar{w}) \lambda \bar{\mu})\;. \end{equation} We consider
$\Omega=\{\rho<0\}$ and $X=\partial\Omega=\{\rho=0\}$. We may assume
without
loss of generality that $h(z,\bar{w}) = \bar{h}(w,\bar{z})$ since $h(z,
\bar{z})$ is real so we could replace $h$ by $\frac{1}{2}h(z,\bar{w}) +
\half\bar{h}(w,\bar{z})$. On $X$ we  have $h(z, \bar{z}) |\lambda|^2 =
1$ so
we may write $\lambda = h(z, \bar{z})^{-\half } e^{i \phi}$, and
similarly
for $\mu$.  So for $(x,y) = (z, \phi, w, \phi') \in X \times X$ we have
\begin{equation} \psi(z, \phi, w, \phi') = \frac{1}{i} \left[1 -
\frac{h(z,\bar{w})}{ \sqrt{h(z, \bar{z})} \sqrt{h(w, \bar{w})}} e^{i
(\phi -
\phi')}\right]\;. \end{equation} It follows that
\begin{equation}\begin{array}{l}\displaystyle\psi( z_0 +
\frac{u}{\sqrt{N}},
\theta; z_0 + \frac{v}{\sqrt{N}}, 0)\\[14pt] \displaystyle \quad\quad=
\frac{1}{i} \left[1 - \frac{h(z_0 + \frac{u}{\sqrt{N}},\bar{z_0} +
\frac{\bar{v}}{\sqrt{N}})}{ \sqrt{h(z_0 + \frac{u}{\sqrt{N}}, \bar{z_0}
+
\frac{\bar u}{\sqrt{N}})} \sqrt{h(z_0 + \frac{v}{\sqrt{N}}, \bar{z_0} +
\frac{\bar v}{\sqrt{N}})}} e^{i \theta}\right] .
\end{array}\end{equation}

We now assume that $e_L$ is a normal frame centered at
$z_0$. By definition, this means that \begin{equation} h(z_0)=1,\quad
\d h |_{z_0} = \dbar h|_{z_0} = 0. \end{equation} We furthermore assume
that
our coordinates $\{z_j\}$ are chosen so that the Levi form of $h$ is
the identity at $z_0$:
\begin{equation}\frac{\d^2 h}{\d z^\al \d z^\be}(z_0,\bar z_0)
=\delta_{\al\be}\,.\end{equation} (This is equivalent to specifying that
$\om(z_0)=\frac{i}{2}\sum_j dz_j \wedge d\bar z_j$.)  Then by
(\ref{taylorh}),
\begin{equation}\label{taylorh2} h(z_0 + \frac{u}{\sqrt{N}},\bar z_0 +
\frac{\bar v}{\sqrt{N}}) = \frac{1}{N}u\cdot \bar v +
O(N^{-3/2}). \end{equation} 

Now let us return to the phase.  It is given by
\begin{equation}\label{phase}
t \left[ 1 - \frac{h(z_0 + \frac{u}{\sqrt{N}},\bar z_0 + \frac{\bar
v}{\sqrt{N}})}{h(z_0 + \frac{u}{\sqrt{N}},\bar z_0 + \frac{\bar
u}{\sqrt{N}})^{\half} h(z_0 + \frac{v}{\sqrt{N}},\bar z_0 + \frac{\bar
v}{\sqrt{N}})^{\half}} e^{i \theta}\right] - i \theta. \end{equation} By
(\ref{taylorh2}), the phase (\ref{phase}) has the form: \begin{equation}
(t[ 1
- e^{i \theta}] - i \theta) + \frac{t}{N}\left[u\cdot\bar v - \half
|u|^2 -
\half |v|^2\right]e^{i \theta} + O(N^{-3/2}). \end{equation} It is now
evident
that $\Pi_N(z_0 + \frac{u}{\sqrt{N}}, 0; z_0 + \frac{v}{\sqrt{N}}, 0)$
is
given by an oscillatory integral with phase $(t [ 1 - e^{i \theta}] - i
\theta)$; 
the latter two terms can be absorbed into the amplitude.

Thus we have: \begin{equation}\begin{array}{l}  \Pi_N(z_0 +
\frac{u}{\sqrt{N}}, 0; z_0 + \frac{v}{\sqrt{N}}, 0)\\[14pt]\quad =
 N \int_0^{\infty} \int_0^{2\pi}
 e^{ i N (t [ 1 - e^{i \theta}] - i \theta)} e^{t[u\cdot\bar v - \half
|u|^2 -
\half |v|^2] +O(N^{-1/2})}  s(z_0 + \frac{u}{\sqrt{N}}, \theta; z_0
+ \frac{v}{\sqrt{N}}, 0), t N) d\theta
dt.\end{array}\end{equation} We may then evaluate the integral
asymptotically by the stationary phase method as in \cite{Z}.  The phase
is
precisely the same as occurs in $\Pi_N(x,x)$, and as discussed in
\cite{Z}, the single critical point occurs at $t = 1, \theta = 0$.  We
may
also Taylor-expand the amplitude to determine its contribution to the
asymptote. 
Precisely as in the calculation of the stationary phase expansion in
\cite{Z}, 
we get: \begin{equation} \Pi_N(z_0 +
\frac{u}{\sqrt{N}}, 0; z_0 + \frac{v}{\sqrt{N}}, 0) = 
\frac{N^m}{\pi^m} e^{u\cdot\bar v - \half |u|^2 -
\half |v|^2} + O(N^{m - \half}). \end{equation} 
Finally, we note that $$u\cdot\bar v - \half |u|^2 -
\half |v|^2=-\half |u-v|^2 +i\Im (u\cdot \bar v)\,,$$ 
which completes the proof
of Theorem \ref{neardiag}.\qed

\subsection{Universality of the scaling limit of correlations of zeros}

We are now ready to pass to the scaling limit as $N\to\infty$ of the
correlation functions of
sections of powers $L^N$ of our line bundle.  To
explain
this notion, let us  consider the case $k=m$ where the zeros are (almost
surely) discrete.  An $m$-tuple of sections of $L^N$ will have $N^m$
times
as many zeros as $m$-tuples of sections of $L$.  Hence we must expand
our
neighborhood (or contract our ``yardstick") by a factor of $N^{m/2}$. 
Let $z^0\in M$ and choose a coordinate neighborhood $U\in M$ with
coordinates
$\{z_j\}$ for which
$z^0=0$ and
$\omega(z^0)=\frac{i}{2}\sum_q dz_q\wedge d\bar z_q$.  We define the
{\it
$n$-point scaling limit zero correlation function\/}
$$K_{nkm}^\infty(z)=\lim_{N\to\infty}\frac{1}{ N^{nk}}
K_{nk}^N\left(\frac{z}{
\sqrtn}\right)\,,\quad z=(z^1,\dots,z^n)\in (\C^m)_n\,.$$  

We show below (Theorem \ref{uslc}) that this limit exists and that
$K_{nkm}^\infty$ is universal by passing to the limit in
Theorem \ref{npointcor}, using Theorem \ref{neardiag}.  First, we need
the
following fact:

\begin{lem}\label{detinfty} Let $z^1,\dots,z^n$ be distinct points
of $\C^m$.  Then
$$\det\left(\Pi^\H_1(z^p,0;z^{p'},0)\right)= e^{-\sum|z^p|^2}\det\left(
e^{z^p\cdot\bar z^{p'}}\right) \ne 0\;.$$\end{lem}

\begin{proof} We consider the first Szeg\"o projector on the reduced
Heisenberg group
\begin{equation} \Pi^\H_1:\lcal^2(\H^m_{\rm red})\to \hcal^2_1(\H^m_{\rm
red})\approx \lcal^2(\C^m,e^{-|z|^2})\cap\ocal(\C^m)\,,\end{equation}
where
$$ 
\lcal^2(\C^m,e^{-|z|^2})=\left\{f\in\lcal^2_{\rm loc}(\C^m):
{\textstyle \int_{\C^m}}|f|^2 e^{-|z|^2}dz <+\infty\right\}\,.$$
(See the remark at the end of \S \ref{example2}.)  Its kernel can be
written
in the form \begin{equation}
\Pi^\H_1(z,\theta;w,\phi)=e^{i(\theta-\phi)}
\sum_{\al=1}^\infty f_\al(z)\overline{f_\al(w)}\,,\end{equation}
where the $f_\al$ form a complete orthonormal basis for 
$\lcal^2(\C^m,e^{-|z|^2})\cap\ocal(\C^m)$.  (E.g., $\{f_\al\}$ can be
taken
to be the set of monomials $\left\{c_{j_1\cdots j_m}z_1^{j_1}\cdots
z_m^{j_m}\right\}$.  In fact, $\Pi^\H_1(z,0;w,0))$ is just a ``weighted
Bergman kernel" on $\C^m$.) We now mimic the proof of Lemma \ref{detN},
except this time we have an infinite sum over the index $\al$; this sum
converges uniformly on bounded sets in $\C^m\times \C^m$ since the sup
norm
over a bounded set is dominated by the Gaussian-weighted $\lcal^2$ norm
(by
the same argument as in the case of the ordinary Bergman kernel on a
bounded
domain).  We then obtain a nonzero vector
$(v_1,\dots,v_n)\in\C^m$ such that
$\sum_p v_p f_\al(z^p)=0$ for all $\al$. But then $\sum_p v_p
f(z^p)=0$ for all polynomials $f$ on $\C^m$, a contradiction.  
\end{proof}

We can now show the universality of the scaling limit of the zero
correlation functions:

\begin{theo}\label{uslc} Let $(L,h)$ be a positive Hermitian line bundle
on
an
$m$-dimensional compact complex manifold $M$ with \kahler form
$\om=\frac{i}{2}\Theta_h$, let
$\scal=H^0(M,L^N)^k$ ($k\ge 1$), and give $\scal$ the standard Gaussian
measure $\mu$.  Then 
$$\frac{1}{ N^{nk}} K_{nk}^N\left(\frac{z^1}{\sqrtn}, \dots,
\frac{z^n}{\sqrtn}\right) = K_{nkm}^\infty(z^1,\dots,z^n) +
O\left(\frac{1}{\sqrtn}\right)\,,$$
where $K_{nkm}^\infty(z^1,\dots,z^n)$ is given by a universal rational
function in the quantities $z^p_q, \bar z^p_q, e^{z^p\cdot \bar
z^{p'}}$,
and the error term has $\ell^{th}$ order derivatives $\le
\frac{C_{S,\ell}}{\sqrtn}$ on each compact subset $S\subset (\C^m)_n$,
for all $\ell\ge 0$. \end{theo} 

\begin{proof} By taking the scaling limit of (\ref{npointcoreq}), 
we obtain 
\begin{equation}\label{uscleq}
K_{nkm}^\infty(z)= \frac{\pcal_{nkm}\big(\Pi^\H_1(z^p,z^{p'}),d^H_{\bar
w_{q}}\Pi^\H_1(z^p,z^{p'}), d^H_{z_{q}}\Pi^\H_1(z^p,z^{p'}), 
d^H_{z_{q}}d^H_{\bar
w_{q'}}\Pi^\H_1(z^p,z^{p'})\big)}{\pi^{kn}\left[\det
\big(\Pi^\H_1(z^p,z^{p'})\big)_{1\le p,p'\le n}\right]^{k(n+1)}}\,.
\end{equation}
Indeed, since the coefficients of $\La_n$ are either of degree 1 in
the coefficients of $C_n$ or of degree 2 in the coefficients of $B_n$,
we see by the proof of Theorem
\ref{npointcor}, using (\ref{dPiNH}), (\ref{dms5})--(\ref{dms6}) and
Theorem \ref{neardiag}, that the leading term of the asymptotic
expansion of
$K_{nk}^N$ is $N^{nk}$ times the right side of (\ref{uscleq}). The bound
on
the error term follows from Theorem \ref{neardiag} and Lemma
\ref{detinfty}. 

Substituting into (\ref{uscleq}) the values of $\Pi^\H_1(z^p,z^{p'})$
and its
horizontal derivatives obtained from (\ref{dPiNHleft}) (with $N=1$) and
(\ref{sksl}) and canceling out common factors of $e^{-|z^p|^2/2}$ and
$\pi$,
we obtain
\begin{eqnarray}
K_{nkm}^\infty(z)&=& \frac{\pcal_{nkm}\big(e^{z^p\cdot \bar
z^{p'}},(z^{p}_q-z^{p'}_q) e^{z^p\cdot \bar z^{p'}}, (\bar z^{p'}_q 
-\bar z^{p}_q)e^{z^p\cdot\bar
z^{p'}},  [(z^p_{q'}-z^{p'}_{q'})(\bar z^{p'}_q-\bar 
z^{p}_q)+\de_{qq'}]e^{z^p\cdot\bar
z^{p'}}\big)}{\pi^{kn}\left[\det
\big(e^{z^p\cdot\bar z^{p'}}\big)_{1\le p,p'\le
n}\right]^{k(n+1)}}\nonumber
\\
\label{uscleq2left}\\
&=& \frac{\qcal_{nkm}\big(z^p_q,\bar z^p_q, e^{z^p\cdot \bar
z^{p'}}\big)}{\pi^{kn}\left[\det
\big(e^{z^p\cdot \bar z^{p'}}\big)\right]^{k(n+1)}}\,,\nonumber
\end{eqnarray} where $\qcal_{nkm}$ is a universal polynomial
(homogeneous of
degree $k(n+1)$ in each of the variables $e^{z^p\cdot \bar z^{p'}}$ and
with
integer coefficients).
\end{proof}

\begin{rem} As we remarked previously, formula (\ref{uscleq}) is valid
for any
connection, so we can replace the left invariant vector fields with
their
right-invariant counterparts to obtain
\begin{equation}\label{uscleq2}
K_{nkm}^\infty(z)= \frac{\pcal_{nkm}\big(e^{z^p\cdot \bar
z^{p'}},z^{p}_q e^{z^p\cdot \bar z^{p'}}, \bar z^{p'}_q e^{z^p\cdot\bar
z^{p'}},  (z^p_{q'}\bar z^{p'}_q+\de_{qq'})e^{z^p\cdot\bar
z^{p'}}\big)}{\pi^{kn}\left[\det
\big(e^{z^p\cdot\bar z^{p'}}\big)_{1\le p,p'\le n}\right]^{k(n+1)}}\,.
\end{equation}\end{rem}

\bigskip
\section {Formulas for the scaling limit zero correlation
function}\label{formulas} 

We now apply the formulas from \S\S
\ref{formula-gaussian}--\ref{D-meet-S} to
transform (\ref{uscleq2}) into explicit formulas for $K_{nkm}^\infty$. 
We use
the right-invariant connection $\alpha^R$ so that $d^H_{z_q} = Z_q^R.$
Indeed,
by the proofs of Theorems \ref{npointcor} and \ref{uslc} (which use
formulas
(\ref{fg6}), (\ref{fg8}), (\ref{dms4})--(\ref{dms6})), formula
(\ref{uscleq2})
becomes
\begin{equation}\label{a28}
K_{nkm}^\infty(z^1,\dots,z^n)=\frac{1}{\pi^{kn}\det A_{nkm}}\left\langle
\prod_{p=1}^n \det (a^pa^{p*})\right\rangle_{\La_{nkm}}\,,
\end{equation}
where
\begin{equation}\label{a29}
\La_{nkm}=C_{nkm}-B^*_{nkm}A^{-1}_{nkm}B_{nkm}
\end{equation}
with
\begin{equation}\label{a30}
\begin{array}{ll}
A_{nkm}=\left(\de_{jj'}S(z^p,z^{p'})\right)\,,&\displaystyle
S(z,w)=\exp\left(\sum_{r=1}^m z_r\overline{w_r}\right),\\
B_{nkm}=\left(\de_{jj'}S_{q'}(z^p,z^{p'})\right)\,,
&\displaystyle
S_{q'}(z,w)
=z_{q'}\exp\left(\sum_{r=1}^m z_r\overline{w_r}\right)\,, \\ 
C_{nkm}=\left(\de_{jj'} S_{qq'}(z^p,z^{p'})\right)\,,&\displaystyle
S_{qq'}(z,w)=(\de_{qq'}+\overline{w_q}z_{q'})
\exp\left(\sum_{r=1}^m z_r\overline{w_r}\right)\\
j,j'=1,\dots,k;\quad p,p'=1,\dots,n;&\displaystyle q,q'=1,\dots,m. 
\end{array}
\end{equation}
The metric tensor $g^p$ in (\ref{fg8}) becomes a unit tensor 
in the scaling limit, so there
is no $\gamma^p$ on the right in (\ref{a28}).

Because $\Pi_1^\H$ is invariant with respect to unitary transformations
and equivariant with respect to translations (i.e.,
$\Pi_1^\H(z+u,w+u)=e^{i\Im(z\cdot \bar u)} e^{-i\Im(w\cdot \bar u)}
\Pi_1^\H(z,w)$), the scaling limit zero correlation $K_{nkm}^\infty$ is
invariant with respect to the group of isometric
transformations---unitary transformations and
translations---of $\C^m$.

In particular, the limit one-point zero correlation, or the zero-density
function, is constant, since it is invariant under translation. Indeed
by
(\ref{a30}), $A_{1km}=e^{|z|^2}I_k$ and $\La_{1km}=e^{|z|^2}I_{km}$,
where
$I_k$, resp.\ $I_{km}$, denotes the unit $k\times k$, resp.\ $(km)\times
(km)$, matrix.  Thus by (\ref{a28}) and the Wick formula,
\begin{equation}\label{a32e} K_{1km}^\infty(z)=\frac{1}{\pi^k
e^{k|z|^2}}\left\langle
\det\left(\sum_{q=1}^m
\bar a_{jq}{a_{j'q}}\right)_{j,j'=1,\dots,k}
\right\rangle_{e^{|z|^2}I_{km}}
=\frac{m!}{\pi^k(m-k)!}\,.\end{equation}
Thus we define the {\it normalized n-point scaling limit zero
correlation
function} \begin{equation}\label{nslzc} \wt K^\infty_{nkm}(z)=
(K^\infty_{1km})^{-n}K^\infty_{nkm}(z)=\left(\frac{\pi^k(m-k)!}{m!}\right)^n
K^\infty_{nkm}(z)\,.\end{equation}

\begin{rem} These formulas also follow from \S \ref{zcorpoly}.  For
example,
equation (\ref{a32e}) is a consequence of (\ref{a32b}) since
$$K_{1km}^\infty(z)=\frac{1}{N^k} K_{1k}^N(z)\,.$$
Furthermore, using the notation of \S \ref{zcorpoly}, we
observe that
\begin{equation}\label{a27}
\begin{array}{ll}
&\displaystyle\lim_{N\to\infty} S_N\left(\frac{z}{\sqrtn},
\frac{w}{\sqrtn}\right)=
\lim_{N\to\infty}\left(1+N^{-1}\sum_{r=1}^m z_r\overline{w_r}\right)^N
=S(z,w)\,,\\
&\displaystyle\lim_{N\to\infty} N^{-1/2}S_{Nq'}\left(\frac{z}{\sqrtn},
\frac{w}{\sqrtn}\right)=
S_{q'}(z,w)\,,\\
&\displaystyle\lim_{N\to\infty} N^{-1}S_{Nqq'}\left(\frac{z}{\sqrtn},
\frac{w}{\sqrtn}\right)=
S_{qq'}(z,w)\,.
\end{array}
\end{equation}
Equations (\ref{a27}) provide an alternate derivation of (\ref{a30}).
\end{rem}

\subsection{Decay of correlations}\label{decay} 
Explicit formulas for the correlation functions $\wt K^\infty_{nkm}$ can be
obtained from (\ref{a28}), (\ref{a30}) and the Wick formula.  We shall
illustrate these computations for the cases $n=2,\ k=1,2$ in \S\S
\ref{explicit1}--\ref{explicit2} below.  We now note that the limit
correlations are ``short range" in the following sense:

\begin{theo}\label{shortrange} The correlation functions satisfy the estimate
$$\wt
K^\infty_{nkm}(z^1,\dots,z^n) = 1 +O(r^4 e^{-r^2}) \quad {\rm as}\ r\to
\infty\,, \quad r=\min_{p\ne p'}|z^p-z^{p'}|\,.$$ \end{theo}

\begin{proof} We use formula (\ref{uscleq}), which comes
from (\ref{a28})--(\ref{a29}) as in the proof of Theorem \ref{uslc}. To
determine the matrices $A,B,C$, we let $d^H_{z_q}=Z^L_q,\ d^H_{\bar
w_q}=\bar W^L_q$ (instead of the right-invariant vector fields we used
above).  Recalling (\ref{dPiNHleft}), we have:

\begin{eqnarray}\label{ABC} A^{jp}_{j'p'}&=& \de_{jj'} A^p_{p'}\,,\qquad
A^p_{p'} = \pi^m \Pi_1^\H(z^p,0;z^{p'},0)\,,\nonumber \\
B^{jp}_{j'p'q'} &=&  \de_{jj'}(z^{p}_{q'}-z^{p'}_{q'}) A^p_{p'}\,,\\
C^{jpq}_{j'p'q'} &=&  \de_{jj'}\big(\de_{qq'}+(\bar z^{p'}_q-\bar z^{p}_{q})
( z^{p}_{q'}-z^{p'}_{q'})\big) A^p_{p'}
\,.\nonumber \end{eqnarray}
By (\ref{sksl}), \begin{eqnarray*} A^p_{p'}&=&\left\{\begin{array}{ll}1 &
p=p'\\  O(e^{-r^2/2})\, \quad & p\ne p'\end{array}\right.\,,\\
B&=&O(re^{-r^2/2})\,,\\
C&=&I+O(r^2
e^{-r^2/2})\,,\qquad C^{jpq}_{jpq}=1
\,.\end{eqnarray*}
Recalling (\ref{fg6}), we have
\begin{equation}\label{Lambda}\La=I+O(r^2 e^{-r^2/2})\,,\qquad
\La^{jpq}_{jpq}=1+O(r^2 e^{-r^2})\,.\end{equation} We now apply formula
(\ref{a28}); note that the Wick formula involves terms that are  products
of diagonal elements of $\La$, and products
that contain at least two off-diagonal elements of $\La$.  The former terms
are of the form $1+O(r^2e^{-r^2})$, and the latter are $O(r^4e^{-r^2})$.
Similarly, $\det A=1+O(r^4e^{-r^2})$. The desired estimate then
follows from (\ref{nslzc}).
\end{proof}

We shall see from our computations of the pair correlation below that Theorem
\ref{shortrange} is sharp.  The theorem can be extended to
estimates of the connected  correlation functions (called also truncated
correlation functions, cluster functions, or cumulants), as follows. The
$n$-point connected correlation function is defined as
(see, e.g., \cite[p.~286]{GJ}) 
\begin{equation}\label{cc1}\wt
T^{\infty}_{nkm}(z^1,\dots,z^n)=\sum_G(-1)^{l+1}(l-1)!
\prod_{j=1}^l \wt K^{\infty}_{n_jkm}(z^{i_1},\dots,z^{i_{n_j}}),
\end{equation}
where the sum is taken over all partitions $G=(G_1,\dots,G_l)$ of the
set $(1,\dots,n)$ and $G_j=(i_1,\dots,i_{n_j})$. In particular, recalling that
$\wt K^\infty_{1km}\equiv 1$,
\begin{eqnarray*}\label{cc2}
\wt T^{\infty}_{1km}(z^1)&=&\wt K^{\infty}_{1km}(z^1)=1\,,\\
\wt T^{\infty}_{2km}(z^1,z^2)&=&\wt K^{\infty}_{2km}(z^1,z^2)
-\wt K^{\infty}_{1km}(z^1)\wt K^{\infty}_{1km}(z^2)
\ =\ \wt K^{\infty}_{2km}(z^1,z^2)-1\,,\\
\wt T^{\infty}_{3km}(z^1,z^2,z^3)&=&\wt K^{\infty}_{3km}(z^1,z^2,z^3)
-\wt K^{\infty}_{2km}(z^1,z^2)\wt K^{\infty}_{1km}(z^3)
-\wt K^{\infty}_{2km}(z^1,z^3)\wt K^{\infty}_{1km}(z^2)\\
&&\ -\ \wt K^{\infty}_{2km}(z^2,z^3)\wt K^{\infty}_{1km}(z^1)
+2 \wt K^{\infty}_{1km}(z^1)\wt K^{\infty}_{1km}(z^2)
\wt K^{\infty}_{1km}(z^3)\\
&=&\wt K^{\infty}_{3km}(z^1,z^2,z^3)
-\wt K^{\infty}_{2km}(z^1,z^2) -\wt K^{\infty}_{2km}(z^1,z^3) 
-\wt K^{\infty}_{2km}(z^2,z^3)+2\,, \end{eqnarray*}
and so on. The inverse of (\ref{cc1}) is
\begin{equation}\label{cc3}\wt
K^{\infty}_{nkm}(z^1,\dots,z^n)=\sum_G
\prod_{j=1}^l \wt T^{\infty}_{n_jkm}(z^{i_1},\dots,z^{i_{n_j}})
\end{equation}
(Moebius' theorem).
The advantage of the connected correlation functions is that they
go to zero if at least one of the distances $|z^i-z^j|$ goes to infinity (see
Corollary \ref{ccc} below).
In our case the connected correlation functions can be estimated as
follows. Define
\begin{equation}\label{cc4}
d(z^1,\dots,z^n)=\max_{\gcal}\prod_{l\in L}
|z^{i(l)}-z^{f(l)}|^2e^{-|z^{i(l)}-z^{f(l)}|^2/2}.
\end{equation}
where the maximum is taken over all oriented connected graphs
$\gcal=(V,L,i,f)$ such that $V=(z^1,\dots,z^n)$ and for every vertex $z^j\in
V$ there exist at least two edges emanating from $z^j$. Here $V$ denotes the
set of vertices of $\gcal$, $L$ the set of edges, and $i(l)$ and $f(l)$ stand
for the initial and final vertices of the edge $l$, respectively. Observe that
the maximum in (\ref{cc4}) is achieved at some graph $\gcal$, because
$te^{-t/2}\le 2/e<1$ and therefore the product in (\ref{cc4}) is less or equal
$(2/e)^{|L|}$ which goes to zero as $|L|\to\infty$.

\begin{theo}\label{cc} The connected correlation functions satisfy the estimate
$$\wt T^\infty_{nkm}(z^1,\dots,z^n) = O(d(z^1,\dots,z^n)) \quad {\rm as}\
\max_{p,q}|z^p-z^q|\to \infty\,,$$ provided that
$\min_{p,q}|z^p-z^q|\ge c>0$. \end{theo}

This theorem implies Theorem \ref{shortrange} because of the inversion formula
(\ref{cc3}). To prove the theorem let us remark that we can rewrite (88)
(using the Wick theorem) as a sum over Feynman diagrams. Namely,
for the normalized correlation functions  $\wt K^\infty_{nkm}(z^1,\dots,z^n)$
we have that
\begin{equation}\label{cc5}
\wt K^\infty_{nkm}(z^1,\dots,z^n)
={[(m-k)!/m!]^n\over \det A_{nkm}}
\sum_{\fcal} A_{\fcal}(z^1,\dots,z^n)\,,
\end{equation}
where the sum is taken over all graphs $\fcal=(V,L,i,f)$ (Feynman diagrams)
such that $V=(z^1,\dots,z^n)$ and the edges $l\in L$ connect the paired
variables $a^{i(l)}_{jq},\; a^{f(l)*}_{j'q'}$ in a given term of the Wick sum
for $\wt K^\infty_{nkm}(z^1,\dots,z^n)$.  The function
$A_{\fcal}(z^1,\dots,z^n)$ is a sum over all terms in the Wick sum with a
fixed Feynman diagram $\fcal$. In other words, to get
$A_{\fcal}(z^1,\dots,z^n)$ we fix pairings $(a^p_{jq},a^{p'*}_{j'q'})$
prescribed by $\fcal$ and sum up in the Wick formula over all indices $j,q$ at
every $a^p$. A remarkable property of the connected correlation functions is
that they are represented by the sum over connected Feynman diagrams (see,
e.g., \cite{GJ}),
\begin{equation}\label{cc6}\wt T^\infty_{nkm}(z^1,\dots,z^n)= 
{[(m-k)!/m!]^n\over \det A_{nkm}}{\sum_{\fcal}}^{\rm conn} 
A_{\fcal}(z^1,\dots,z^n)\,.\end{equation}
Since $\det A_{nkm}\ge c_1>0$ and $|\La^{jpq}_{j'p'q'}|\le c_2<+\infty$ when
$\min_{p,q}|z^p-z^q|\ge c>0$, we conclude from (\ref{fg6}), (\ref{ABC}) and
(\ref{sksl}) that for all connected Feynman diagrams $\fcal$,
\begin{equation}\label{cc7}
A_{\fcal}(z^1,\dots,z^n)=O(d)\,,
\end{equation}
where $d=d(z^1,\dots,z^n)$ is defined in (\ref{cc4}).
Summing up over $\fcal$, we prove Theorem \ref{cc}.\qed

\begin{cor}\label{ccc} The connected correlation functions satisfy the estimate
$$\wt T^\infty_{nkm}(z^1,\dots,z^n) = O(R^2e^{-R^2/2}) \quad {\rm as}\
R\to\infty\,,\quad R=
\max_{p,q}|z^p-z^q|\,,$$ provided that
$\min_{p,q}|z^p-z^q|\ge c>0$. \end{cor}

\subsection {Hypersurface pair correlation}\label{explicit1} We now give
an
explicit formula [(\ref{a40})] for pair correlations in codimension 1
($k=1,\ n=2$).  The case $m=1$ of this formula coincides, as it must,
with
the formula given by
\cite{H} and \cite{BBL} for the universal scaling
limit pair correlation for $\SU(2)$ polynomials. In
another paper
\cite{BSZ}, we gave a different proof of (\ref{a40}) using the
Poincar\'e-Lelong formula.

Since the  scaling-limit pair correlation function
$K_{2km}^\infty(z^1,z^2)$ is invariant with respect to the group of
isometries
of
$\C^m$, it depends only on the distance
$r=|z^1-z^2|$, so we can
set $z^1=0$ and
$z^2=(r,0,\dots,0)$. To simplify
notation, we shall henceforth write $A=A_{2km},\; B=B_{2km},\;
C=C_{2km},\;
\La=\La_{2km}$. 

In this case, (\ref{a30}) reduces to
\begin{equation}\label{a33}
\begin{array}{lll}
A=
\left(\begin{array}{ll}
1 & 1 \\
1 & e^{r^2}
\end{array}\right);\\
B=(B^p_{p'q});&\ 
(B^p_{p'1})=
\left(\begin{array}{ll}
0 & 0 \\
r & re^{r^2}
\end{array}\right);&\quad 
(B^p_{p'q})=
\left(\begin{array}{ll}
0 & 0 \\
0 & 0
\end{array}\right),\quad q\ge 2;\\
C=(C^{pq}_{p'q'});&\ (C^{p1}_{p'1})=
\left(\begin{array}{ll}
1 & 1 \\
1 & (1+r^2)e^{r^2}
\end{array}\right);
&\quad (C^{pq}_{p'q'})=\de_{qq'}
\left(\begin{array}{ll}
1 & 1 \\
1 & e^{r^2}
\end{array}\right),\quad q,q'\ge 2.
\end{array}
\end{equation}
The matrix 
\begin{equation}\label{a34}
\La=(\La^{pq}_{p'q'})=C-B^*A^{-1}B
\end{equation}
is given by
\begin{equation}\label{a35}
\La^{p1}_{p'1}=
\left(\begin{array}{ll}
\di\frac{e^u-1-u}{e^u-1} & \di\frac{e^u-1-ue^u}{e^u-1} \\
\di\frac{e^u-1-ue^u}{e^u-1} & \di\frac{e^{2u}-e^u-ue^u}{e^u-1}
\end{array}\right);\quad
\La^{pq}_{p'q'}=\de_{qq'}
\left(\begin{array}{ll}
1 & 1 \\
1 & e^u
\end{array}\right),\ q,q'\ge 2\,,
\end{equation} where  $u=r^2=|z^1-z^2|^2$.
By (\ref{a28}), (\ref{nslzc}) and the formula for $A$ in (\ref{a33}), we
have
\begin{equation}\label{a36}
\wt K^\infty_{21m}(z^1,z^2)=\frac{1}{m^2(e^u-1)}\left\langle 
\left(\sum_{q=1}^m a^1_q\overline{a^1_q}\right)
\left(\sum_{q'=1}^m
a^2_{q'}\overline{a^2_{q'}}\right)\right\rangle_{\La}
\end{equation}
By the Wick formula (see for example, \cite[(I.13)]{Si}),
\begin{equation}\label{a37}\begin{array}{lll}
\wt K^\infty_{21m}(z^1,z^2)&=&\di\frac{1}{m^2(e^u-1)}\left[
\left(\sum_{q=1}^m\langle  a^1_q\overline{a^1_q}\rangle_{\La_2}\right)
\left(\sum_{q'=1}^m\langle 
a^2_{q'}\overline{a^2_{q'}}\rangle_{\La_2}\right)
+\sum_{q,q'=1}^m \langle  a^1_q\overline{a^2_{q'}}\rangle_{\La_2}
\langle  \overline{a^1_q}a^2_{q'}\rangle_{\La_2}\right]\\[14pt]
&=&\di\frac{1}{m^2(e^u-1)}\left[
\left(\sum_{q=1}^m \La^{1q}_{1q}\right)\left(\sum_{q'=1}^m
\La^{2q'}_{2q'}\right)+\sum_{q,q'=1}^m \La^{1q}_{2q'}
\La^{2q'}_{1q}\right]\,.\end{array}\end{equation}
Substituting the values of $\La^{pq}_{p'q'}$
given by (\ref{a35}), we obtain
\begin{equation}\label{a37a}
\begin{array}{lll}
\wt K^\infty_{21m}(z^1,z^2)&=&\di\frac{1}{m^2(e^u-1)}\left[
\left(\frac{e^u-1-u}{e^u-1}+m-1\right)
\left(\frac{e^{2u}-e^u-ue^u}{e^u-1}+(m-1)e^u\right)\right.\\[14pt]
&&\quad\di\left.+\left(\frac{e^u-1-ue^u}{e^u-1}\right)^2+(m-1)\right]\,,
\qquad u=|z^1-z^2|^2\,.\end{array}
\end{equation}
After simplification,
\begin{equation}
\wt K^\infty_{21m}(z^1,z^2)=\frac{u^2(e^{2u}+e^u)-2u(e^{2u}-e^u)
+m^2(e^u-1)^2e^u+m(e^u-1)^2}{m^2(e^u-1)^3}\,.
\end{equation}
Putting $u=2t$ and writing
\begin{equation}\label{a39}
\wt
K^\infty_{21m}(z^1,z^2)=\kappa_{1m}(|z^1-z^2|)\,,
\end{equation}
we then obtain
\begin{equation}\label{a40}
\kappa_{1m}(r)=\frac{\left[\frac{1}{2}(m^2+m)\sinh^2t+t^2\right]
\cosh t
-(m+1)t\sinh t}{m^2\sinh^3t}+\frac{(m-1)}{2m},\quad
t=\frac{r^2}{2}\,.
\end{equation}
The case $m=1$ of formula (\ref{a40}) was obtained by
Bogomolny-Bohigas-Leboeuf \cite{BBL} and Hannay \cite{H}.  

As $r\to\infty$,
\begin{equation}\label{a40a} \kappa_{1m}(r)=1 +
\frac{r^4-2(m^2+1)r^2+m(3m+1)}{m^2}e^{-r^2} +O(r^4e^{-2r^2})\,.
\end{equation}
The following expansion of
the correlation function was obtained from (\ref{a40}) using Maple$^{\rm
TM}$:
\begin{eqnarray*} 
\kappa_{1m}& =&\frac{m-1}{2m}t^{-1}+\frac{m-1}{2m}+
\frac{1}{6}\,{\frac {\left (m+2\right )\left (m+1\right )}{{m}^{2}}}t
-{\frac {1}{90}}\,{\frac {\left (m+4\right )\left (m+3\right )}
{{m}^{2}}}{t}^{3}\\&&\ +
{\frac {1}{945}}\,{\frac {\left (m+6\right )\left (m+5\right )}
{{m}^{2}}}{t}^{5}
-{\frac {1}{9450}}\,{\frac {\left (m+8\right )\left (m+7\right )
}{{m}^{2}}}{t}^{7}\\&&+
{\frac {1}{93555}}\,{\frac {\left (m+10\right )\left (m+9\right 
)}{{m}^{2}}}{t}^{9}
-{\frac {691}{638512875}}\,{\frac {\left (m+12\right )\left (m+11
\right )}{{m}^{2}}}{t}^{11}\\&&+
{\frac {2}{18243225}}\,{\frac {\left (m+14\right )\left (m+13\right 
)}{{m}^{2}}}{t}^{13}
-\cdots\;.
\end{eqnarray*}
In particular, in the one-dimensional case we have
\begin{equation}\label{a40b} \kappa_{11}(r)=\frac{1}{2}r^2 -
\frac{1}{36}r^6
+\frac{1}{720}r^{10} -\frac{1}{16800}r^{14} +  \cdots\;.\end{equation}

\subsection{Pair correlation in higher codimension}\label{explicit2}
Next we compute the
two-point correlation functions for the case
$k=2$. For $k>1$, we have
\begin{equation}\label{b1} A=(A^{jp}_{j'p'})=(\de_{jj'}A^p_{p'})\,,\quad
B=(B^{jp}_{j'p'q'})=(\de_{jj'}B^{p}_{p'q'})\,,\quad
C=(C^{jpq}_{j'p'q'})=(\de_{jj'}C^{pq}_{p'q'})\,,\end{equation}
where $A^p_{p'},B^{p}_{p'q'},C^{pq}_{p'q'}$ are given by (\ref{a33}).
It
follows that 
\begin{equation}\label{b2}
\La=(\La^{jpq}_{j'p'q'})=(\de_{jj'}\La^{pq}_{p'q'})\,,\end{equation}
where $\La^{pq}_{p'q'}$ is 
given by (\ref{a35}).

By (\ref{a28}),
\begin{equation}\label{a41}
K^\infty_{2km}(z^1,z^2)=\frac{1}{\pi^{2k}(e^u-1)^k}\left\langle 
\det\left|a_j^1\overline{a_{j'}^1}\right|_{j,j'=1,\dots,k}
\det\left|a_j^2\overline{a_{j'}^2}\right|_{j,j'=1,\dots,k}
\right\rangle_{\La},\quad
a_j^p\overline{a_{j'}^{p}}=\sum_{q=1}^m
a^p_{jq}\overline{a^{p}_{j'q}}\,,
\end{equation}
where $u=r^2=|z^1-z^2|^2$ as before.
Observe that the random variables $a^p_{jq}$ and
$\overline{a^{p'}_{j'q'}}$ are 
independent if either $j\not= j'$ or $q\not= q'$. 

Recalling (\ref{nslzc}), we write
\begin{equation}
\wt K^\infty_{2km}(z^1,z^2) =\kappa_{km}(|z^1-z^2|)\,.
\end{equation}
When $k=2$, (\ref{a41}) reduces to the
following 
\begin{equation}\label{a42}
\kappa_{2m}(r)=\frac
{\left\langle\left[ (a^1_1\overline{a^1_1})( a^1_2\overline{a^1_2}) 
-(a^1_1\overline{a^1_2})( a^1_2\overline{a^1_1})\right]
\left[ (a^2_1\overline{a^2_1})( a^2_2\overline{a^2_2})
-(a^2_1\overline{a^2_2})( a^2_2\overline{a^2_1})\right]
\right\rangle_{\La}}{m^2(m-1)^2(e^u-1)^2}\,.
\end{equation}
By the Wick formula,
\begin{equation}
\kappa_{2m}(r)=\frac{d_{11}-d_{21}-d_{12}+d_{22}}{
m^2(m-1)^2(e^u-1)^2}\,,
\end{equation} where
\begin{equation}\label{a43}
\begin{array}{rl}
d_{11}= &\di\left\langle (a^1_1\overline{a^1_1})(
a^1_2\overline{a^1_2})
 (a^2_1\overline{a^2_1})( a^2_2\overline{a^2_2})
\right\rangle_{\La}
=\sum_{\al,\be,\ga,\de}\left\langle a^1_{1\al}\overline{a^1_{1\al}} 
a^1_{2\be}\overline{a^1_{2\be}} a^2_{1\ga}\overline{ a^2_{1\ga}}
a^2_{2\ga}\overline{ a^2_{2\ga}}
\right\rangle_\La\\
=&\di \sum_{\al,\be,\ga,\de}
\La^{1\al}_{1\al}\La^{1\be}_{1\be}
\La^{2\ga}_{2\ga}\La^{2\de}_{2\de}
\ + 2\sum_{\al,\be,\ga}
\La^{1\al}_{1\al}\La^{1\be}_{2\be}
\La^{2\be}_{1\be}\La^{2\ga}_{2\ga}
\ + \sum_{\al,\be}
\La^{1\al}_{2\al}\La^{2\al}_{1\al}
\La^{1\be}_{2\be}\La^{2\be}_{1\be}\\
=&\left[\left(\sum_{q}\La^{1q}_{1q}\right)
\left(\sum_{q}\La^{2q}_{2q}\right)
\ + \sum_q \La^{1q}_{2q}\La^{2q}_{1q}\right]^2\,;
\end{array}
\end{equation}
similarly,
\begin{equation}\label{a43a}
\begin{array}{rl} d_{12}= &\di
\left\langle (a^1_1\overline{a^1_1})(
 a^1_2\overline{a^1_2}) 
 (a^2_1\overline{a^2_2})( a^2_2\overline{a^2_1})
\right\rangle_{\La}\\
=&\left(\sum_q \left[\La^{2q}_{2q}\right]^2\right)\left(\sum_q
\La^{1q}_{1q}\right)^2  +
2\left(\sum_q\La^{2q}_{2q}\La^{2q}_{1q}\La^{1q}_{2q}\right)
\left(\sum_q\La^{1q}_{2q}\right) +\sum_q
\left[\La^{2q}_{1q}\La^{1q}_{1q}\right]^2\,,
\end{array}
\end{equation}
\begin{equation}\label{a43b}
\begin{array}{rl}
d_{21}= &\di
\left\langle (a^1_1\overline{a^1_2})( a^1_2\overline{a^1_1})
 (a^2_1\overline{a^2_1})( a^2_2\overline{a^2_2})
\right\rangle_{\La}\\
=&\left(\sum_q \left[\La^{1q}_{1q}\right]^2\right)\left(\sum_q
\La^{2q}_{2q}\right)^2  +
2\left(\sum_q\La^{1q}_{1q}\La^{2q}_{1q}\La^{1q}_{2q}\right)
\left(\sum_q\La^{2q}_{2q}\right) +\sum_q
\left[\La^{2q}_{1q}\La^{1q}_{2q}\right]^2\,,
\end{array}
\end{equation}
\begin{equation}\label{a43c}
\begin{array}{rl}
d_{22}= &\di
\left\langle (a^1_1\overline{a^1_2})( a^1_2\overline{a^1_1})
(a^2_1\overline{a^2_2})( a^2_2\overline{a^2_1})
\right\rangle_{\La}\\ 
=&\left(\sum_q \left[\La^{1q}_{1q}\right]^2\right)\left(\sum_q
\left[\La^{2q}_{2q}\right]^2\right)  +
2\sum_q\La^{1q}_{1q}\La^{2q}_{2q}\La^{2q}_{1q}\La^{1q}_{2q}
+\left(\sum_q
\La^{2q}_{1q}\La^{1q}_{2q}\right)^2\,. 
\end{array}
\end{equation}

Substituting the values of the matrix elements of $\La$ we then
obtain
\begin{equation}
\begin{array}{ll}
\kappa_{2m}(r)&= \di
\frac{(m^2-m)e^{2u}+2(m-1)e^u+2}{(e^u-1)^2m(m-1)}
-\frac{4ue^u[(m-1)e^u+1](m+1)}{(e^u-1)^3(m-1)m^2}\\
&\\
&\di +\frac {2u^2e^u[(m-1)e^{2u}+2me^u+1]}{(e^u-1)^4(m-1)m^2},\qquad
u=r^2. 
\end{array}
\end{equation}

As $r\to\infty$,
\begin{equation}\label{z}
\kappa_{2m}=
1+\frac{2[r^4-2(m+1)r^2+m(m+1)]e^{-r^2}}{m^2} +O(r^4e^{-2r^2})\,.
\end{equation} 
As $r\to 0$,
\begin{equation}
\begin{array}{rl}
\kappa_{2m}(r)&\di =\frac{m-2}{m}\,r^{-4}+\frac{m-2}{m}\,r^{-2}
+\frac{5m^2-7m+12}{12(m-1)m}
+\frac{(m-2)(m+2)(m+1)}{12(m-1)m^2}\,r^2\\
&\\
&\di +\frac{(m+3)(m+2)}{240(m-1)m}\,r^4
-\frac{(m-2)(m+4)(m+3)}{720(m-1)m^2}\,r^6+\dots
\end{array}
\end{equation}
When $m=2$ the asymptotics reduce to
\begin{equation}
\kappa_{22}(r)=\frac{3}{4}+\frac{r^4}{24}-\frac{r^8}{288}
+\frac{r^{12}}{4800}-\frac{r^{16}}{96768}+\dots,
\end{equation}
and in this case $\kappa_{22}$ is a series in $r^4$.

\end{document}